\documentclass[aps,pre,reprint,groupedaddress,showpacs,showkeys]{revtex4-1}

\usepackage{bm}
\usepackage{epsfig}
\usepackage{graphicx,graphics}
\usepackage{amsmath}
\usepackage{amsfonts}
\usepackage{amssymb}
\usepackage{epstopdf}

\begin{document}

\title{Statistical mechanics of thermal fluctuations of  nearly spherical membranes: the influence of bending and stretching elasticities}

% repeat the \author .. \affiliation  etc. as needed
% \email, \thanks, \homepage, \altaffiliation all apply to the current
% author. Explanatory text should go in the []'s, actual e-mail
% address or url should go in the {}'s for \email and \homepage.
% Please use the appropriate macro foreach each type of information

% \affiliation command applies to all authors since the last
% \affiliation command. The \affiliation command should follow the
% other information
% \affiliation can be followed by \email, \homepage, \thanks as well.

%\email[]{bivas@issp.bas.bg}
\author{\bf{Nicholay S. Tonchev}}

\email[]{tonchev@issp.bas.bg}
%\homepage[]{http://...}
%\thanks{aaa}
%\altaffiliation{aaa}
\affiliation{Institute of Solid State Physics, Bulgarian Academy of Sciences,
	72~Tzarigradsko chaussee blvd., Sofia~1784, Bulgaria, }

	\vspace{0.80 cm}

	\vspace*{20pt}
	\begin{abstract}
		\begin{flushright}
			{\it To the memory of my friend and colleague Slava Priezzhev, an\\ outstanding scientist and man, whose contribution to statistical\\ mechanics  and  probability theory  is original and remarkable in \\ many ways. His comments on physics and things of life were\\ always nontrivial and impactful.}
\end{flushright}
		
\begin{center}		
{\bf Abstract:}
\end{center}

Theoretical studies of nearly spherical vesicles and microemulsion droplets, that present   typical examples for 
thermally-excited systems that are  subject to constraints,  are reviewed. 
We consider the  shape  fluctuations of such   systems constrained by fixed area $A$ and fixed volume $V$,  whose geometry is presented  in terms of scalar spherical harmonics. These constraints   can be  incorporated in the theory in  different ways.
After an introductory review of the two approaches: with an exactly fixed by delta-function membrane area $A$ [Seifert,  Z. Phys. B,  97, 299, (1995)]  or
approximatively by means
of a Lagrange multiplier
$\sigma$ conjugated to $A$ [Milner and Safran, Phys. Rev. A,  36,	4371 (1987)], 
we discuss the 
determined role of the stretching effects, that has been announced in the framework of a model containing stretching energy term, expressed via the membrane vesicle tension [Bivas and Tonchev, Phys.Rev.E, 100, 022416 (2019)]. Since the fluctuation spectrum for the used Hamiltonian   is not exactly solvable an approximating method based on the Bogoliubov inequalities for the free energy has been developed. The area constraint in the last approach appears as a self-consistent equation for the membrane tension. In the general case this equation is intractable analytically. However, much insight into the physics behind can be obtained  either imposing some restrictions on the values of the model  parameters, or  studying limiting cases, in which  the self-consistent equation is solved. Implications for the equivalence of ensembles have been  discussed as well.
\end{abstract}
\vspace*{6pt}
\onecolumngrid

\noindent
%PACS:87.16.D$-
%\onecolumngrid
%\begin{widetext}
	%\begin{align}
	
%\end{widetext}
\label{sec:intro}
	\maketitle
\section*{Introduction}

The biomembrane consists of a lipid bilayer, in which
integral proteins float~\cite{Sin72}. Within this model, their physical properties
are tightly connected with those of the lipid bilayer.
The complicated role of biological membranes helped forward the study of artificial lipid-based models with a primary view of reconstituting their inherent functions {\it in vitro}. Lipid membranes are important model systems for biological membranes, see \cite{P99} and refs. therein.  Easily formed from lipid solutions  this field involves the investigation of transport mechanisms, permeation properties, adhesion, and fusion kinetics, see \cite{Am17} and refs. therein.

In this review we shall consider the role of thermal fluctuations on the behavior of closed free standing artificial membranes. To a large extent, the physics of closed membranes is the physics of microemulsions and vesicles (also known as liposomes).  

The microemulsions are thermodynamically stable water-oil mixtures and one or few surface active agents (called also surfactants). Probably one of the simplest 
microemulsion systems are {\it droplets} of water in oil (or vice versa), covered with a monolayer of surfactant molecules.

The vesicles are analogous to droplet type microemulsions  where the most important difference is that their interface is
a bilayer instead of a monolayer.
They are closed-surface  membranes formed spontaneously from molecules in aqueous environment due to the hydrophobic effect. For example,
Giant unilamellar vesicles (GUVs) have diameters in the interval 1 - 200 $\mu m $. Because of the extremely small thickness of the membrane as compared to the square root of their area, the vesicle for many purposes can 
be modeled as {\it  two-dimensional flexible surface}  embedded in three-dimensional space. Thought both the vesicle and the microemulsion systems belong to quite different length scale, their thermodynamic behavior  can be understood from a unified point of view based on study of the free energy of the deformed  membrane \cite{S03}. 

The  shape of a monolayer membrane depends mainly on its stretching  $g_s$ and
curvature $g_c$ energy densities
\begin{equation}
g=g_s+g_c.
\label{a12}
\end{equation}
The elastic energy of stretching per unit area (in a quadratic approximation) has the form
\begin{equation}
g_s=\frac{1}{2}K_s\Bigg(\frac{\Delta A }{A_0}\Bigg)^2,
\end{equation}
where $\Delta A/A_0$ is the relative area change of the membrane, $A_0$ is the area in its tension free state, and $K_s$ is the stretching elasticity modulus \cite{H73}. 

The curvature elasticity, as one of the most important quantities for determining  the  membrane shape, was introduced in the beginning of the seventies of the last century  with the theory developed by  Canhamm \cite{C70}, Helfrich \cite{H73} and Evans \cite{E74}. 
The theory of spherical vesicles is based on the notion that their 
bending elastic energy  is size independent, since the bending energy per unit area of a symmetric bilayer is a quadratic form in  curvatures \cite{H86}.
In the used quadratic approximation the curvature-elastic
energy per unit area of a fluid layer may be written as \cite{H73}
\begin{equation}
g_c=\frac{1}{2}K_c(c_1+c_2-c_s)^2 + K_Gc_1c_2
\label{g11}
\end{equation}
Here $c_1$, $c_2$  are the principal curvatures and $c_s$ is the
spontaneous curvature. The
constant $K_c$ is the bending rigidity modulus. 
The last term in \eqref{g11} contains   the modulus of the 
Gaussian curvature $K_G$, which 
should be  omitted, in general, as when integrated over a closed
surface it is a topologically invariant constant. 
The derivation of $K_s$ and $K_c$, from the first principles  for the membrane energy functional $g$, is given in \cite{MSWD94}; where   it has been  obtained for the typical phospholipids, $K_s\sim 200$ erg/cm$^2$ and
$K_c \sim 2\times 10^{-12}$ erg.
Actually, the quadratic form Eq.~\eqref{g11} seems to be the best possible approximation consistent with isotropy, fluidity and  Euclidean invariance shape dependence of the membrane (for a comment, see Appendix A of ref.\cite{MSWD94}). For more complete and detailed comments the reader can see the review \cite{D15}, where the tools of differential geometry are used to obtain the Helfich Hamiltonian.

The effective Hamiltonian (the total bending energy stored in the infinitely  thin interface $A$ of the vesicle in a second order expansion of curvatures) has the form:
\begin{equation}
H_c=\int_A dS \Bigg[\frac{1}{2}K_c(c_1+c_2-c_s)^2 + K_Gc_1c_2\Bigg]
\label{g1a1}
\end{equation}
where the integration is carried out over the total area $A$ of the membrane.

At first in our consideration, the membranes
will be assumed to be unstretchable (viz. with $K_s=\infty$) 
and their finite
thickness will  be disregarded. As we shall see below,  the former is not a obligatory requirement, while  the last approximation is assumed in order to avoid   difficulties  of integration over random surfaces \cite{L04} in the calculations of the free energy and corresponding thermodynamic mean values.
The material quantities $K_s$ and $K_c$ depend on the intermolecular interactions and structural properties of the membrane. Elasticity  of lipid monolayers in the framework of different  molecular models
are discussed in some details in Ch. 4.2 of the monograph \cite{P99}.
Strictly speaking, $K_s$ and $K_c$ cannot be considered independently of the thermal
fluctuations, because integrating out the microscopic degrees of freedom, in the Helfrich theory, the resulting coarse-grained Hamiltonian~\eqref{g1a1} will contain the temperature dependent phenomenological parameters $K_s$ and $K_c$. The convention in this review is that when we talk  about thermal fluctuations we take into account only  fluctuations of the geometry of the membrane \cite{D15}.
The spectrum of thermally excited
undulation modes of a quasi-spherical vesicle can be calculated analytically in the limit of large bending rigidities $K_c/k_BT>>1$, $T$ is the temperature and $k_B$ is the Boltzmann constant (Helfrich 1986 \cite{H86},
Milner and Safran 1987 \cite{MS87}, Seifert 1995 \cite{S95}). It can be applied to determine
experimentally the bending rigidity $K_c$ through the  
the contour fluctuations of quasi-spherical vesicles
using phase contrast microscopy combined with fast image processing, nowadays also  called  as flicker spectroscopy analysis 
\cite{BL75,SJW84,B87,F89,Mel97,PDPJB04,GVB13,G13,VP13,MS16,R17}.
\label{sec:preparation}
%\section*{Paper Preparation}
%\end{document}
\section {Membrane geometry  and the Helfrich bending Hamiltonian in terms of spherical harmonics.}

We consider a  vesicle with volume $V$ that has the form of slightly deformed sphere, i.e. a quasi-sphere. Let us denoted by $R_0$  the radius of an ideal spherical vesicle with the same volume $V=(4\pi/3)R_0^3$ as that the studied one. The statistical fluctuations   of the shape of the vesicle membrane in time are around the reference sphere with area $4\pi R_0^2$.
In this case the membrane configuration can be well represented using spherical harmonic functions.
Let us take the quasi-sphere radius
${\cal R}$  to be a function of the time $t$, 
the polar angle $\phi$ and the azimuthal
angle $\theta$, writing
\begin{equation}
{\cal R}(\theta,\phi, t)=R_0[1 + u(\theta,\phi,t)],
\label{r1}
\end{equation}
thus $R_0=[3/(4\pi)V]^{1/3} $ is the effective radius 
from which the displacements (the fluctuations)
$u(\theta,\phi,t)$ are supposed to start.  It is assumed that $|u(\theta,\phi,t)|<<1$, so that all the further calculations are performed up to the second order in $u(\theta,\phi,t)$ in order to simplified the theoretical expressions.
The dimensionless function $u(\theta, \varphi,t)$ can be
decomposed
in a series of spherical harmonics as follows:
%\cite{H86,MS87,S95}:
\begin{equation}
u(\theta,\varphi,t) =
\sum_{n=0}^{n_{max}}\sum_{m=-n}^{n}
u_n^m(t) Y_n^m(\theta,\varphi), \label{e6}
\end{equation}
where $Y_n^m(\theta,\varphi)$ is the orthonormal basis 
of the spherical harmonics functions  and since the displacement is real the relation $(u^m_n)^*=(-1)^m u^{-m}_n$ takes holds. 
For vesicles of slightly deformed spherical shapes, one can calculate the bending
energy $H_c$ by inserting the spherical harmonics expansion Eq.~\eqref{e6} into
Eq.~\eqref{g1a1} and use, as a rule, the lowest approximation neglecting all terms of higher than quadratic order in coefficients
$u_n^m$.
It can  be proved that  all using  expressions containing complex amplitudes may be rewritten in such a way that the imaginary parts
of $u_n^m$ vanish and only real amplitudes appear. Thus without loss of generality $u^m_n$  should be chosen
real. For details see, e.g. \cite{H86,B87,Oh89,H92,S10}. For numerical computations it is also more convenient to use real amplitudes of spherical harmonics instead of the complex ones.

The lipid vesicle is subject to various geometric constraints.      If it is viewed as impermeable and incompressible the number of  lipid molecules in the membrane is fixed. In this case
the vesicle has constrained area $A$ and constrained volume $V$. 
Expressing the geometrical quantities $A$ and $V$, and the effective Hamiltonian $H_c(\mathrm{u})$ of the vesicle as a function of the expansion coefficients $u^m_n$
one has \cite{H86,MS87,F89,Oh89,H92,KS93,S95,Bif10,GDM17,S94}:
\begin{eqnarray}
&&A(\mathrm{u})=4\pi R_0^2 + 4\pi R_0^2\left[\frac{u_0^0}{\sqrt{\pi}} + \frac{ (u_0^0)^2}{4\pi}\right] +\nonumber\\
&&R_0^2\sum_{n\ge
	 1}^{n_{max}}\sum_{m=-n}^{n}\left[1+\frac{n(n+1)}{2}\right](u^m_n)^2 +O((u^m_n)^4),
\label{gS}
\end{eqnarray}
for the area, and
\begin{eqnarray}
&&V(\mathrm{u})=\frac{4}{3}\pi R_0^3\times\nonumber\\
&&\left[\left(1+\frac{u^0_0}{\sqrt{4\pi}}\right)^3\right. + 
\left.\frac{3}{4\pi}\sum_{n\ge 1}^{n_{max}}\sum_{m=-n}^{n}(u^m_n)^2\right] +O((u^m_n)^3),\nonumber\\
\label{gV}
\end{eqnarray}
for the volume. Here and hereafter the symbol  $"\mathrm{u}" $ is used as a
shorthand for the real value  functions $(u_2^{-2}, u_2^{-1},\dots,
u_{n_{max}}^{n_{max}})$, which are the spherical harmonics amplitudes, appearing in
the expansion of the vesicle shape fluctuations from the equivalent volume sphere with radius $R_0$ (see Eq.~\eqref{e6}).

A cut-off $n_{max}\sim
2\sqrt{\pi} R_0/\lambda $ is introduced in the sum,  where $\lambda$ is of the order of the intermolecular distance.  As the harmonics with indexes $n=1$ and $m=-1,0,1$
correspond to pure translation of the vesicle, the origin $O$ of the coordinate system can be chosen in away that $u_1^m=0$. Thus, from now on  all sums start at $n=2$.

Finally, 
considering  the special case of vanishing spontaneous
curvature ($c_s = 0$), for the effective Hamiltonian  $H_c(\mathrm{u})$ in terms of $u^m_n$, one finds:

\begin{eqnarray}
&&H_c(\mathrm{u})
=8\pi K_{c}+\nonumber\\
&&\frac{K_c}{2}\sum_{n\ge 2}^{n_{max}}\sum_{m=-n}^{n}(n-1)n(n+1)(n+2)(u^m_n)^2 +O((u^m_n)^3)\nonumber\\
\label{he11}
\end{eqnarray}
Further we shall discard in $H_c(\mathrm{u})$ the constant energy term $8\pi\overline{k}$.
%\end{document}
\section{The area and the volume constraints} 

The constant-area constraint or the constant-volume constraint can be easily incorporated in the theory, separately \cite{KS93,S95,Gk96}, by choosing $u^0_0$ to satisfy $A=4\pi R_0^2$ in Eq.\eqref{gS}, or $V=(4/3)\pi R_0^3$ in Eq.\eqref{gV}.
However, if it is technically  easy to implement the second constraint, then it is difficult to handle the first one or vice-versa. Regardless of
either volume or area is kept constant these conditions leads to the elimination of the $u^0_0$-term.
In what follows, we shall assumed the vesicle  volume $V$ to be invariant under shape fluctuations.
Thus, the volume constraint 
\begin{equation}
V(\mathrm{u})\equiv \frac{4}{3}\pi R_0^3 
\label{cR}
\end{equation}
implies 
\begin{equation}
u^0_0=-\sqrt{\frac{1}{4\pi}}\sum_{n \ge 2}^{n_{max}}\sum_{m=-n}^{n}(u^m_n)^2 + O((u^m_n)^3).
\label{vc}
\end{equation}
After that, inserting Eq.~\eqref{vc} in Eq.~\eqref{gS}, for the area, keeping terms to $O(u^2)$, one gets
\begin{equation}
A(\mathrm{u})=4\pi R_0^2\left[1 +  \frac{1}{8\pi}\sum_{n \ge 2
}^{n_{max}}\sum_{m=-n}^{n}(n+2)(n-1)(u^m_n)^2\right].
\label{gS1}
\end{equation}
%where the subscript $V$ denotes that the volume is fixed.
Now, it is a more sophisticated  problem  to fix 
the area, Eq.~\eqref{gS1}, of the membrane:
\begin{equation}
A(\mathrm{u})= A.
\label{cA}
\end{equation}
In order to include Eq.~\eqref{cA} in our consideration one needs to use the methods of statistical mechanics. 
First of all let us recall some basic definitions. We start from the Helmholtz free
energy $f[H(\mathrm{u})]$:
\begin{equation}
f[H(\mathrm{u})]=-k_BT\ln\{Z[H(\mathrm{u})]\}, \label{e60}
\end{equation}
where $Z[H(\mathrm{u})]$ is the partition function of the considered quasi-spherical membrane (droplet microemulsion and vesicle) with an effective Hamiltonian $H(\mathrm{u})$ (see below):
\begin{equation}
Z[H(\mathrm{u})]  =\int{\it D}\{\mathrm{u}\}
\{\exp{[-\beta H(\mathrm{u})]}\}, \label{e61}
\end{equation}
where $\beta= (k_BT)^{-1}$ is the inverse temperature .

Note that the correct definition of the measure
${\it D}\{\mathrm{u}\}$ is a subtle task in statistical mechanics of two-
dimensional surfaces (see \cite{S95,S94,GDM17,CLNP94,D04} and refs. therein). However, for a quasi-spherical membrane we don't need to go beyond the so-called normal gauge, which is known to be correct for small fluctuations, at least to the level of accuracy consistent with the used quadratic approximation in Eq.~\eqref{he11}\cite{S95,S94,Bif10}. At this level the proper measure is ${\it D}\{\mathrm{u}\}=const.(d[u_2^{-2}], d[u_2^{-1}],\dots,
d[u_{n_{max}}^{n_{max}}])$ and the integration over $u_{n}^{m}$ in Eq.~\eqref{e61} is carried out from $0$ to $\infty$.

If the Hamiltonian $H(\mathrm{u})$ is a positive definite diagonal quadratic form in  the  real value  functions $(u_2^{-2}, u_2^{-1},\dots,
u_{n_{max}}^{n_{max}})$ then the  multiple integral in Eq.~\eqref{e61} over   $u^m_n$
splits into a product of $N=\sum_{n=2}^{n=n_{max}}\sum_{m=-n}^{m=n}=n_{max}^2+2n_{max}-3$  one-dimensional  Gaussian integrals and the integration can be performed easily.

Commonly there are two alternative possibilities  how to incorporate the area constraint in the partition function Eq.~\eqref{e61}: exact treatment of the area constraint with a delta function \cite{Bi92,S95,S94,Fap01} and  with an effective tension through   Lagrange multiplier \cite{MS87,F89,S95,S97}.  	It is evident that the two approaches
model two different statistical ensembles; $A$-ensemble and $\sigma$-ensemble.The equivalence of the ensembles is taken for granted
too often in  the membrane fluctuation theories.
%\cite{F11,Hnf16}.
However, it is questionable  whether this equivalence holds for all  characteristics of the system.  As we show below this problem needs to be carefully
considered. 
%\end{document}

\section{The exact treatment of the area constraint}

First, in our consideration we keep the global surface area constant, so we can assume that the area compressibility modulus $K_s=\infty$. The exact treatment of the area constraint by delta-function is based on the evaluation of the following partition function (see,e.g.\cite{S95,Fap01}):
\begin{equation}
Z[H_c(\mathrm{u});A]  =\int{\it D}\{\mathrm{u}\}\delta\left(\frac{A-A(\mathrm{u})}{4\pi R_0^2}\right)
\exp[-\beta H_c(\mathrm{u})]. \label{e61c}
\end{equation}
Hereafter for convenience the dependence on $A$ in the argument of the delta-function is explicitly shown in the arguments of the partition function  and in the corresponding free energy 
\begin{equation}
F[H_c(\mathrm{u});A]=-\beta^{-1}\ln\{Z[H_c(\mathrm{u});A]\} \label{e60c}.
\end{equation}
In order to impose condition Eq.\eqref{cA} we use the Laplace transform representation of the delta-function
\begin{equation}
\delta(x-y)=\frac{1}{2\pi i}\int_{a-i\infty}^{a+i\infty}e^{s(x-y)}ds,
\end{equation}
where $s$ is a complex variable and its real part is $a>0$.
In our case we have (compare with Eq.~(15) in \cite{Fap01})
%\begin{aligh}
\begin{widetext}
\begin{equation}
\delta	\left(\frac {1}{8\pi}\sum_{n\ge 2}^{n_{max}}\sum_{m=-n}^{n}(n+2)(n-1)(u^m_n)^2-\Delta\right)
=
\frac{1}{2\pi i}
\int_{a-i\infty}^{a+i\infty}ds 
\exp\left[s\left(\frac{1}{8\pi}\sum_{n\ge 2}^{n_{max}}\sum_{m=-n}^{n}(n+2)(n-1)(u^m_n)^2-
\Delta\right)\right],
\label{dl}
\end{equation}
\end{widetext}
%\end{aligh}
where
\begin{equation}
\Delta\equiv\frac{A-4\pi R^2_0}{4\pi R^2_0},
\end{equation}
defines the dimensionless (positive) excess area used in the further consideration as a small parameter in the model. Note that our definition of the excess area differs from those used in \cite{S95} by $4\pi$. After interchanging  the integrals in Eq.~\eqref{e61c}, one obtains
\begin{equation}
Z[H_c(\mathrm{u});A]  =\frac{1}{2\pi i}
\int_{a-i\infty}^{a+i\infty}ds e^{s} Z[H(\mathrm{u};s)].
\label{e61d}
\end{equation}
Here, $Z[H(\mathrm{u};s)]$ is the partition function of the temperature dependent  "Hamiltonian"
%\begin{widetext}
	\begin{eqnarray}
&&H(\mathrm{u};s)= H_c(\mathrm{u})+ \nonumber\\
&&\frac {s}{\beta} \left[\frac{1}{8\pi}\sum_{n\ge 2}^{n_{max}}\sum_{m=-n}^{n}(n+2)(n-1)(u^m_n)^2 - \Delta\right],
\label{GH1}
\end{eqnarray}
%\end{widetext}
with the  auxiliary complex parameter $s$. The real part $a>0$ of $s$
is chosen so that the integral in Eq.~\eqref{e61d} is finite.
It will be convenient to introduce a new dimensionless quantity
\begin{equation}
\overline{\sigma}_{s}= \frac{s}{4\pi\beta K_c}.
\end{equation}
From Eqs.~\eqref{GH1}  and \eqref{he11}, we obtain:
\begin{equation}
H(\mathrm{u},s)= 4\pi K_c\Delta\overline{\sigma}_{s} + \sum_{n=2}^{n_{max}}\sum_{m=-n}^{n}a_n(\overline{\sigma}_{s}) (u_n^m)^2,
\label{AH1d}  
\end{equation}
where
\begin{equation}
a_n(\overline{\sigma}_{s})=\frac{1}{2} K_c(n-1)(n+2)\left[n(n+1)+\overline{\sigma}_{s}\right].
\label{AH2d}
\end{equation}

Now, it is convenient  to introduce the following quantities:

\begin{eqnarray}
&&w^m_n\equiv u^m_n\left(\frac{K_c\beta}{2}(n+2)(n-1)\right)^{1/2},\nonumber\\
&& p_n\equiv n(n+1),\quad \frac{1}{\tau}\equiv 4\pi K_c\beta\Delta.
\end{eqnarray}
Then partition function $Z[H(\mathrm{u};s)$ is presented as:
\begin{eqnarray}
&&\frac{Z[H(\mathrm{u};s)}{ Z_0}=\nonumber\\
&&\int D\{\mathrm{w}\}\exp\Bigg(- \sum_{n\ge 2}^{n_{max}}\sum_{m=-n}^{n}(p_n +{\overline\sigma}_s)(w^m_n)^2+
\frac{s}{\tau}\Bigg),\nonumber\\
\end{eqnarray}
where $Z_0$ is an independent of $s$ factor. 
The partition function $Z[H(\mathrm{u};s)]$ is known (the corresponding integrals over $w^m_n$ are Gaussian). Thus one obtains
\begin{equation}
Z[H_c(\mathrm{u});A] \sim \int_{a-i\infty}^{a+i\infty}ds e^{s/\tau}\prod_{n\ge 2}^{n_{max}}(p_n + {\overline\sigma_s})^{-(n+1/2)}.
\label{BS}
\end{equation}

For $\tau <<1$, i.e. $1/(\beta K_c) <<1$ and $\tau >>1$, i.e. $1/(\beta K_c ) >>1$, analytical calculations for the mean square amplitudes $u^m_n$ were performed in refs.~\cite{S95,S94}. For completeness let us briefly sketch the results for each of the two cases.
The integral in Eq.\eqref{BS} can be treated by the method of steepest descent. 

In the case, $\tau<<1$, the results are:
\begin{eqnarray}
&&\langle (u^m_n)^2\rangle_{H(\mathrm{u};A)}=\frac{1}{5}\frac{1}{\beta K_c\tau}\Bigg[\frac{1}{2}-\nonumber\\
&&\tau\sum_{n\ge 3}\frac{1}{(n+2)(n-1)(n^2+n-6)}
+O(\tau^3)\Bigg], 
\label{Se2}
\end{eqnarray}
for $n=2$, and
\begin{eqnarray}
&&\langle (u^m_n)^2\rangle_{H(\mathrm{u};A)}=\frac{1}{\beta K_c} \Bigg[\frac{1}{(n+2)(n-1)(n^2+n-6)}+\nonumber\\
&&O(\tau)\Bigg], 
\label{Se1}
\end{eqnarray}
for $n\ge 3$.
An expansion beyond the leading terms for small $\tau$ is possible but in this case the truncated  higher order terms in the expansions Eqs.~\eqref{gS}~-~\eqref{he11} should be taken into account for the consistency of the used approximation.

In the opposite case, $\tau>>1$, the bending energy can be treated as small perturbation and in lowest order the result is
\begin{equation}
\langle (u^m_n)^2\rangle_{H(\mathrm{u};A)}\approx\frac{1 }{\beta K_c\tau}\Bigg[\frac{2}{N(n+2)(n-1)}\Bigg],
\label{Se1a}
\end{equation}
where $N=\sum_{n\ge 2}^{n_{max}}\sum_{m=-n}^{n}=(n_{max}+1)^2-4$ is the available  number of modes.

%\end{document}

\section{The conventional approach with effective tension}

The  approach to the problem, due to the Milner and Safran (1987), and Seifert (1995), is to treat $Z[H(\mathrm{u};s)]$ in Eq.~\eqref{e61d} as a "grand-canonical" partition function, where now $s$ is a real variable (called {\it effective tension} and denoted with $\sigma$)
conjugate to $ A(\mathrm{u})$, rather than a free parameter. 
The value of $\sigma$ is chosen  so that the area constraint Eq.~\eqref{cA} is satisfied on the average. Note that the situation here strongly resembles the relation between the Berlin and Kac spherical model and Lewis and Wannier mean spherical model in the theory of magnetism, which belong to different ensembles with
"canonical" and "grand canonical" partition functions, see e.g. \cite{J72} and Ch.3 in \cite{B00}.

In other words, instead of working with fixed area $A(\mathrm{u})$, an effective tension as Lagrange
multiplier $\sigma$ conjugated to $A(\mathrm{u})$ has been
used. In this case  the Hamiltonian of the model takes the form:

\begin{equation}
H(\mathrm{u};\sigma)= H_c(\mathrm{u})+\sigma A(\mathrm{u}).
\label{HG}
\end{equation} 
This model is essentially  a Gaussian model with a constraint on the area of the vesicle enforced by $\sigma$. The partition function can be thought  as corresponding to an assemble which
extends
the $A$-ensemble by allowing fluctuations of the area in a system of fixed volume $V$.
The partition function is given by 
\begin{equation}
Z[H(\mathrm{u});\sigma]  =\int{\it D}\{\mathrm{u}\}
\exp{[-\beta H(\mathrm{u};\sigma)]}, \label{e61b}
\end{equation}
and the corresponding free energy is given by
\begin{equation}
F[H(\mathrm{u};\sigma)]=-\beta^{-1}\ln\{Z[H(\mathrm{u});\sigma]\}. \label{e60c1}
\end{equation}
We shall define the thermodynamic average of the quantity $A(\mathrm{u})$ with $H(\mathrm{u};\sigma)$ in the standard way:
\begin{eqnarray}
&&\langle A({\mathrm{u}}) \rangle_{H(\mathrm{u};\sigma)}=\nonumber\\
&&\left\{Z[{H(\mathrm{u};\sigma)}]\right\}^{-1}\int {\it D} \{\mathrm{u}\}	
A(\mathrm{u})\exp{\left[-\beta{H(\mathrm{u};\sigma)}\right]}.
\label{Gmv}
\end{eqnarray}
Let us introduce
\begin{equation}
\overline{\sigma}_{MS}=\frac{R_0^2}{K_c}\sigma
\end{equation}
which is  the Milner and Safran dimensionless effective tension.
If for convenience we change the notation $H(\mathrm{u};\sigma)\equiv H(\mathrm{u};\overline{\sigma}_{MS})$), with the help of Eqs. \eqref{he11} and \eqref{gS1} it is easy to obtain the explicit form of
the Hamiltonian Eq.~\eqref{HG}:
\begin{equation}
H(\mathrm{u};\overline{\sigma}_{MS})
=4\pi{\overline \sigma}_{MS}K_c + \sum_{n\ge 2}^{n_{max}}\sum_{m=-n}^{n}a_n({\overline\sigma}_{MS})(u^m_n)^2,
\label{he1}
\end{equation}
where
\begin{equation}
a_n({\overline\sigma}_{MS})=\frac{1}{2} K_c(n-1)(n+2)\left[n(n+1)+{\overline\sigma}_{MS}\right].
\label{AH2e}
\end{equation}

It is constructive to compare the Hamiltonians Eqs.~\eqref{AH1d} and \eqref{he1}.
The only difference is the change of  ${\overline \sigma}_s$   in the former
with ${\overline \sigma}_{MS}$ in the last.

Here and further on, we accept the convention the bar over any quantity to mean "dimensionless due to the multiplier $R^2_0/K_c$".  

Now, the calculation of the Gaussian integrals in Eqs. \eqref{e61b} and \eqref{Gmv} is  straightforward. 
The area constraint is
\begin{equation}
\langle A(\mathrm{u})\rangle_{H(\mathrm{u};{\overline \sigma}_{MS}))}= A.
\label{AcA}
\end{equation}
or equivalently (compare with Eqs. \eqref{gS1} and \eqref{cA})
\begin{widetext}
\begin{equation}
1=\frac{4\pi R_0^2}{A}\left(1 - \frac{1}{4\pi}\sum_{n \ge 2
}^{n_{max}}\sum_{m=-n}^{n}\left[\frac{(n+2)(n-1)}{2}\right]\left\langle(u^m_n)^2\right\rangle_{H(\mathrm{u};{\overline \sigma}_{MS}))}\right).
\label{gS11}
\end{equation}
\end{widetext}

Thus, the value of ${\overline \sigma}_{MS}$ is determined from the condition that the area constraint Eq.~\eqref{cA} is satisfied in average. A possible equivalence of this new introduced statistical ensemble with
those considered in the previous Section 3 we shall discus in the next Section 5.

Here, for the mean square amplitudes 
\begin{eqnarray}
&&\langle (u^m_n)^2 \rangle_{H(\mathrm{u};{\overline \sigma}_{MS})}=\nonumber\\
&&\left\{Z[{H(\mathrm{u};{\overline \sigma}_{MS})}]\right\}^{-1}\int {\it D} \{\mathrm{u}\}	
(u^m_n)^2\exp{\left[-\beta{H(\mathrm{u};{\overline \sigma}_{MS})}\right]}.\nonumber\\
\label{msa}
\end{eqnarray}
one immediately obtains (as follows from Eq.~\eqref{he1} the integrals are Gaussian) 
\begin{equation}
\left\langle(u^m_n)^2\right\rangle_{H(\mathrm{u};{\overline \sigma}_{MS})}=\frac{8\pi \gamma}{(n-1)(n+2)[n(n+1)+{\overline \sigma}_{MS}]},
\label{MSf}
\end{equation}
where
\begin{equation}
\gamma \equiv \frac{1}{8\pi}\frac{1}{\beta K_c}.
\label{g1}
\end{equation}

It is commonly accepted, as a more simple approach, the membrane area constraint to be guaranteed by the Lagrange
multiplier $\sigma$ (see, Eq.\eqref{HG}) conjugate to the real area $A(\mathrm{u})$ \cite{S94,S97,Bif10,F11,Hnf16}. The Lagrange  multiplier  cannot be measured directly, and is experimentally determined thorough the temperature and the physically meaningful quantity named excess area $\Delta_{MS}$ \cite{S95,S97}.
% Contrary to the apparent simplicity of this issue,  the relation of %$\overline{\sigma}_{MS}$ (or $\sigma$) to the other   generic  %definitions
%of membrane surface tension is  a matter of a longstanding debate %(see e.g.\cite{Bif10,F11,Hnf16} and refs. therein). 
The formula Eq.~\eqref{MSf} allows to infer the values of $K_c$ (as follows from Eq.~\eqref{g1}) and $\overline{\sigma}_{MS}$, treating them as fit parameters, from flicker spectroscopy analysis.

By definition the (dimensionless) excess area $\Delta_{MS}$  is related to  a vesicle with fixed volume $V=\frac{4\pi}{3}R^3_0$, and
with  {\it fixed in mean} area, which fluctuates (due to thermal fluctuations
at temperature $T>0$) around the shape $4\pi R^2_0$, through the relation:
\begin{equation}
\Delta_{MS} \equiv \frac{\langle A(\mathrm{u})
\rangle_{H(\mathrm{u};\overline{\sigma}_{MS})}-4\pi R^2_0}{4\pi R^2_0}>0.
\label{a1}
\end{equation}
Combining  Eq.~\eqref{gS11} 
with Eq.~\eqref{a1}, one gets:
\begin{equation}
\Delta_{MS}= \frac{1}{8\pi}\sum_{n \ge 2
}^{n_{max}}\sum_{m=-n}^{n}(n+2)(n-1)\langle(u^m_n)^2
\rangle_{H(\mathrm{u};\overline{\sigma}_{MS})}.
\label{exr}
\end{equation}
Finally, from Eq.~\eqref{MSf} and Eq.~\eqref{exr} one may conclude  that the excess area  obeys the implicit equation \cite{Bi92,S95}
\begin{equation}
\frac{\Delta_{MS}}{\gamma}= 
\sum_{n=2}^{n_{max}}\frac{2n+1}{n(n+1)+
\overline{\sigma}_{MS}(\Delta_{MS})} \label{b101}.
\end{equation}
Thus, after the elimination of the Lagrange multiplier $\overline{\sigma}_{MS}$,
the mean value of the square of the amplitudes $(u^m_n)$, as it follows from Eqs.\eqref{MSf} and \eqref{b101},  depends only on $\Delta_{MS}/\gamma$ and $n_{max}$.

Eq.~\eqref{b101} has been obtained and analyzed by many authors \cite{MS87,Bi92,S95,Bif10,GDM17}.
Excess area $\Delta_{MS}$ as a function of the ratio $\overline{\sigma}_{MS}/N$ for various values of $N$ has been analyzed numerically in ref.\cite{GDM17}.
In such a type of theory (conventional approach with effective tension) the dimensionless  excess area is used  as a small parameter $\Delta_{MS} << 1$. 
Consequently the  term in the rhs of Eq.(\ref{b101}) is also small  and respectively $	\overline{\sigma}_{MS}(\Delta_{MS})$ should be large. 
Our inspection of the Eq.(\ref{b101}) (see the details in \cite{BT19}) shows that the following  functional dependences take place:
\begin{equation}
\overline{\sigma}_{MS}(\Delta_{MS})\approx N\exp\left(-\frac{\Delta_{MS}}{\gamma}\right),
\qquad \exp\left(-\frac{\Delta_{MS}}{\gamma}\right)<<1 
\label{sol02},
\end{equation}
and its inverse  
\begin{equation}
\Delta_{MS}(\overline{\sigma}_{MS})\approx\gamma \ln\left(\frac{N}{\overline{\sigma}_{MS}}
\right),\qquad \frac{N}{\overline{\sigma}_{MS}}>>1,
\label{so202}
\end{equation}
where $N\approx(n_{max})^2$ is the number of lipid molecules in
the vesicle membrane. Further on, when it does not cause confusion, we shall omit the arguments in  $\overline{\sigma}_{MS}$ and $\Delta_{MS}$. 
A similar relation was obtained in ref.\cite{S95} (see also the comment in \cite{S97}) simply by replacing the sum  in the rhs of Eq.~\eqref{b101} with an integral. In our consideration \cite{BT19}  this sum is estimated by
the Euler - McLaurin summation formula. This defines a different range of validity of our
result  given by the inequality in
Eq.~\eqref{sol02}.

As $K_c \to 0$, the bending energy becomes irrelevant for the fluctuation amplitudes, and one may assume  that each mode contributes equally to the excess area\cite{MS87}. Thus, from Eqs.~\eqref{gS11}, and~\eqref{a1} immediately follows
\begin{equation}
\Delta_{MS}=\frac{1}{4\pi}N\left[\frac{(n+2)(n-1)}{2}\right]\left\langle(u^m_n)^2\right\rangle_{H(\mathrm{u};{\overline \sigma}_{MS}))}
\label{gS21}
\end{equation}
in full consistence with Eq.~\eqref{Se1a} obtained above in the exact delta-function approach.
In this limit, from Eq.~\eqref{MSf} one getts
\begin{equation}
\left\langle(u^m_n)^2\right\rangle_{H(\mathrm{u};{\overline \sigma}_{MS})}\approx \frac{8\pi \gamma}{(n-1)(n+2){\overline \sigma}_{MS}}.
\label{MSf1}
\end{equation}
Eliminating $\left\langle (u^m_n)^2\right\rangle_{H(\mathrm{u};{\overline \sigma}_{MS})}$ from both equations, one obtains
\begin{equation}
{\overline \sigma}_{MS}\approx\frac{\gamma}{\Delta_{MS}}N.
\label{sol03}
\end{equation}
Thus, for a vesicle close to a spherical shape, i.e.~$\Delta_{MS}\to 0$, the effective tension is expected
to be proportional to $N$ \cite{MS87,S10}.
Analogous to Eqs.~\eqref{sol02} and~\eqref{sol03} results were derived for almost planar membrane in \cite{Fap01}.
A similar equation to Eq.~\eqref{sol02} was  analyzed  for almost planar membrane in the low-tension regime in ref.~\cite{Hi04} attributed  to entropic-tense and stretched-tense regimes, respectively. The analysis of the
simulation data in the experimentally accessible tension range shows that
the stretching effects of the membrane area must be taken into account.

%\end{document}

\section{Equivalence of  ensembles}
In our case, on physical grounds, one is interested in the situation where the total area  $A$ 
of the vesicle is kept fixed.
Let us mention at once that, instead of working with exactly fixed membrane area $A$ via delta-function, an effective tension as Lagrange
multiplier $\sigma$ conjugated to $A$  has been commonly
used.
In this different $(\sigma;V)$ - statistical
ensemble, the free energy  $F$ depends on $(\sigma;V)$, while in the former 
free energy $f$ depends on $(A;V)$.
From statistical point of view the model with delta-function is defined by the joint probability distribution of the $u^m_n$-variables given by a measure which is different from the corresponding probability distribution  for the model with Lagrange multiplier $\sigma$. 
Thus the two models are in general, statistically  inequivalent. 
Although the two approaches
model two different statistical ensembles, it is widely assumed
that they give equivalent results. 
First of all, it is worth to note that the equivalence of the ensembles is closely related with the  notion of  thermodynamic limit which must be well defined. Thus it will be useful to make a short comment on the subject.

The proof of the equivalence of two statistical ensembles goes back to Gibbs \cite{G02} and is a key problem of the equilibrium statistical mechanics. 
On the contemporary understanding the term {\it equivalence} has three different meanings, each on a different level of information (see, \cite{A06,T15}).
Since the parameter $\sigma$ cannot be measured directly one may think that the most appropriate way  should be the use of the $(A,V)$ - ensemble, 
but as shown in the previous sections  calculations within the $(\sigma,V)$ - ensemble are simpler than the calculations in the $(A,V)$ - ensemble.
Thus in constrained by calculating difficulties one prefers to use the $(A,\sigma)$ and after that goes to the physically relevant parameters $(A,V)$. 
The change from  $(A,V)$ - to $(\sigma, V)$-variables is to be performed via the  Legendre-Fenchel transform, which expresses the thermodynamic potential $f(A,V)$, in terms of $F(\sigma,V)$ as:
\begin{equation}
f(A,V) =  \max_{\sigma}[F(\sigma,V)+\sigma A].
\label{lft}
\end{equation}
It reduces to the Legendre transform in the case of convex, differentiable functions. The first level of ensemble equivalence (i.e. at the level of thermodynamic potentials) takes place if {\it all}  thermodynamic potentials in the {\it thermodynamic limit} are related to each other by a Legendre - transform for the corresponding  values of parameters entering in the Hamiltonian of the system. Roughly speaking this kind of equivalence is usually given in the absence of phase transitions, i.e., only for those thermodynamic parameters where there are no singularities.

The transform Eq.~\eqref{lft} is well behaved only if $F(\sigma,V)$ is a convex function: 
\begin{equation}
\frac{\partial^2 F(\sigma,V)}{\partial^2\sigma} < 0.
\label{2}
\end{equation}
Since  in our case $F(\sigma,V) = F[H(\mathrm{u};\sigma)] $ is a differentiable function of the parameter $\sigma$ Eq.~\eqref{2} holds. It attains its maximum with respect to the later at the unique solution of the equation 
\begin{equation}
\frac{d F(\sigma,V)}{d\sigma}=A,
\end{equation}
or equivalently in more detail
\begin{equation}
\frac{\partial}{\partial \sigma}F[H(\mathrm{u};\sigma)]\equiv-\frac{R_0^2}{\beta K_c}\frac{\partial}{\partial{\overline \sigma}_{MS}}\ln Z[H(\mathrm{u};{\overline \sigma}_{MS})] = A,
\label{ltc}
\end{equation}
which relates implicitly $\sigma$ to $A$. 
Eq.~\eqref{ltc} coincides exactly  with  Eq.\eqref{AcA}.
The thermodynamic of the vesicle  in the $(A,V)$-variable is fully specified by Eqs. \eqref{e60c1} and \eqref{lft}.

More precisely, in the thermodynamic limit, the corresponding canonical free energy density
\begin{equation}
f_\infty(A,V)=\lim_{N\to \infty} \frac{f({A,V})}{N}
\end{equation}
is related to the grand canonical free energy 
\begin{equation}
F_\infty(\sigma,V)=\lim_{N\to \infty} \frac{F({\sigma,V})}{N}
\end{equation}
by means of the Legendre transform,
\begin{equation}
f_\infty(A,V) =  F_{\infty}(\sigma(A),V)+\sigma(A) A
\label{inf}
\end{equation}
where $\sigma(A)$ is the solution of 
\begin{equation}
\frac{d F_{\infty}(\sigma,V)}{d\sigma}=A.
\end{equation}
This is the way to derive the $(A,V)$ - result from the $(\sigma, V)$- one.
Equivalence of both ensembles  holds only if $F_\infty(\sigma,V)$ also can be recovered from $f_\infty(A,V)$ by means of a Legendre transform. The requirement  both type of Legendre transformations
to take place makes the check of equivalence a complicated task solved only in the thermodynamic limit,  i.e. the ensembles are equivalent in the limit of infinitely large membranes.

This is a crucial point which must to be scrutinized in each considered case. As it was pointed out in \cite{Sch11}  there is a situation in Monge model presentation where the ensembles
are not equivalent. The amplitude of fluctuations depend on the
statistical ensemble under consideration as a result of the
ambiguity of the very definition of the thermodynamic limit which compromise the used assumption that the membrane is planar.

On the pure physical ground, there is one more sophisticated  obstacle, namely, infinitely large membrane at finite tension
is an  object not at thermodynamic equilibrium \cite{Sch13}.
This contradicts the basic assumption that the system we considered is assumed to be in thermodynamic equilibrium.

%\end{document}

\section{The stretching elasticity and the soft area constraint}

It is well recognized that the bending modulus $K_c$ is preferably studied   in the literature of the flexibility of membranes (see e.g. the comments in \cite{N13}).
Physical situation in which the membrane may fluctuate through stretching or compressing are of significant interest as well \cite{Bgp76,D97,Hi04,Lli11,Hnf16,BT19,B02,N13, B10, Fp03}.

The presented in this section approach is a reminiscence of an earlier work of Shapiro and Rudnick \cite{Sr86} concerning the quite other field of magnetism. In this work, the spherical constraint (an analog of the area constraint in our case)  is replaced with a Gaussian damping term into the partition function. Here, our present consideration is twofold. First, to trace out an useful further resemblance   with the spherical  model, and second, to introduce in a different way 
a mean {\it softened} area constrained instead  of those enforced with Lagrange  multiplier. 
%We shall discuss below the last scenario so that  to keep track with the %former ones.

An instructive question is whether it is possible to reveal  the microscopic origin of $ {\overline\sigma}_{MS} $ in the formulas of Milner and Safran for the mean square amplitudes Eq.~\eqref{MSf} and excess area Eq.~\eqref{b101}, or more precisely, of the quantities which appear instead. In particular, this should allow to include  the experimental  determination of the stretching elasticity modulus $K_s$ in the scenario of the flicker spectroscopy analysis.

First, let us recall that there is
yet another form of the  delta-function  as the limit of a normalized Gaussian:
\begin{equation}
\delta(x)=\lim_{\epsilon \to +0}\frac{1}{\sqrt{2\pi \epsilon}}\exp(-x^2/2\epsilon).
\label{delta}
\end{equation} 
For our aim, the crucial issue is to define $\epsilon$, which physically must correspond to $K_s$ and $A$, and in addition, if we want to mimic a part of a Hamiltonian in the Gibbs exponential, with the inverse temperature $\beta$.
The only dimensionless combination between $K_s$, $A$ and $\beta$ is $\beta A K_s$. 

In order to make relation with the theory of Milner and Safran we are interested to study the limit $K_s\to \infty$ ($\beta$ and $A$ fixed). The other interesting possibility is to consider the case  $A \to \infty$ ($K_s$ and $\beta$ fixed ). Thus, it is instructive to  choose  the limit:
\begin{equation}
\epsilon:=\frac{1}{\beta A K_s}\equiv \frac{1}{{\cal N}} \to +0
\end{equation}
in the definition Eq.~\eqref{delta}.
Having this in mind, we start with the Gaussian
\begin{equation}
\left(\frac{\beta A K_s}{2\pi}\right)^{1/2}\exp\left\{-\frac{\beta A K_s}{2}\left[\frac{A(\mathrm{u})}{A}-1\right]^{2}\right\}
\label{sc}
\end{equation}
as a factor in the integrand of the partition function $Z[H_c(\mathrm{u})]$. In the limit $\epsilon=0$
this factor becomes a delta function and the exact area constraint \eqref{dl} is recovered 
\begin{equation}
Z[H_c(\mathrm{u});K_s] \stackrel{\cal {N}\to \infty}{\longrightarrow} Z[H_c(\mathrm{u});A],
\label{gra} 
\end{equation}
where, the new (with the relaxed area constraint)  partition function  $Z[H_c(\mathrm{u});K_s]$ is defined as:
\begin{widetext}
\begin{eqnarray}
Z[H_c(\mathrm{u};K_s)]  = 
\left(\frac{\beta A K_s}{2\pi}\right)^{-1/2}\int{\it D}\{\mathrm{u}\}\exp\left\{-\frac{\beta A K_s}{2}\left[\frac{A(\mathrm{u})}{A}-1\right]^{2}\right\}
\exp{\Bigg[-\beta H_c(\mathrm{u})}\Bigg]. \label{e62d}
\end{eqnarray}
\end{widetext}
Also,  as follows  from Eq.~\eqref{e62d},  the  quantity
in the exponentials of $Z[H_c(\mathrm{u});K_s]$, formally may be considered as a Hamiltonian. In this way we come to the effective Hamiltonian:
\begin{equation}
H(\mathrm{u};K_s)=H_c(\mathrm{u})+\frac{K_s}{2 A}[A(\mathrm{u})-A]^{2},
\label{BTH}
\end{equation} 
where the last term in the above expression has the form of the stretching energy functional, 
\begin{equation}
\frac{1}{2}\frac{ [\sigma(\mathrm{u})]^2}{K_s}A =H_s(\mathrm{u})
\label{sef}
\end{equation}
if we express  the membrane vesicle tension 
$\sigma(\mathrm{u})$ as,
\begin{equation}
\sigma(\mathrm{u}) = K_{s}\frac{A(\mathrm{u})-A}{A} . \label{r17}
\end{equation}
% ~\cite{Helf73,Bgp76,Fp03}  and \cite{Hi04}.
The interesting  point here is that we obtain the expression  Eq.\eqref{sef}, based on the requirement of the relaxation of  the exact delta-function  constraint through normalized Gaussian, Eq.~\eqref{delta}. Indeed, Eq.~\eqref{r17} is a definition of $K_s$ which depends on $A$.
Since the stretching modulus $K_s$ can be measured experimentally, in addition, one needs to refine   the meaning of $A$ in order to make track with the experiment \cite{Fap01,Hi04}. We shall postpone the consideration of this issue to the next section.

Now, we shall consider the model Hamiltonian
$H(\mathrm{u};K_s)$  which can be rewritten (up to an irrelevant constant) in the following alternative form:
\begin{equation}
H(\mathrm{u};K_s)={\cal T}(\mathrm{u})+[{\cal A}(\mathrm{u})]^2, \label{e141}
\end{equation}
where
\begin{equation}
{\cal T}(\mathrm{u})=\frac{1}{2} K_c
\sum_{n=2}^{n_{max}}\sum_{m=-n}^n(n-1)(n+2)[n(n+1)+\overline{\sigma}_0](u_n^m)^2,
\label{e140}
\end{equation}
and
\begin{equation}
{\cal A}(\mathrm{u}) =\sqrt{\frac{K_s}{2A}} R_0^2\sum_{n \ge 2
}^{n_{max}}\sum_{m=-n}^{n}\left[\frac{(n+2)(n-1)}{2}\right](u^m_n)^2, \label{e39}
\end{equation}
In Eq.~\eqref{e140}
\begin{equation}
\overline{\sigma}_0=\frac{R^2_0}{K_c}\sigma_0,
\label{e16}
\end{equation}
where the notation 
\begin{equation}
\sigma_0=K_s\frac{4\pi R^2_0-A}{A}. \label{e166}
\end{equation}
has been used.

The considered partition function reads:
%\begin{widetext}
\begin{eqnarray}
&&Z[H(\mathrm{u}; K_s)]:=\nonumber\\
&&\left(\frac{\beta A K_s}{2\pi}\right)^{1/2}\int{\it D}\{\mathrm{u}\}
\exp\Bigg\{-\beta\Bigg[{\cal T}(\mathrm{u})+[{\cal A}(\mathrm{u})]^2\Bigg]\Bigg\}.\nonumber\\
\label{nlt}
\end{eqnarray}
%\end{widetext}
However, then the corresponding Hamiltonian $H(\mathrm{u};K_s)$ in the exponential of Eq.~\eqref{nlt} is nonlinear with respect to the squares
$(u_n^m)^2$ of the amplitudes $u_n^m$, due to the nonlinearity of $H_s(\mathrm{u})$. To solve this
problem one can follow  two different approaches, named: steepest descent method and variational method (see below).  

The common approach is based on the Habbard-Stratonovich transformation (which in our case is nothing but an identity on Gaussian integrals),
with the subsequent use of the steepest descent method
\cite{Fap01,Hi04,Lli11}. It turns out that
the problem
is  exactly solvable (only) in the thermodynamic limit
\cite{Fap01,Hi04,Lli11,F11}. Let us recall that this aspect of the membrane fluctuation theory was first discussed for almost planar membranes in the context of  the spherical model of phase transitions
in 1976 \cite{Bgp76}.

Quite recently, a different approach to linearize the Hamiltonian in Eq.~\eqref{nlt} 
based on the Bogoliubov  variational inequalities has been proposed \cite{BT19}.	
In our opinion this yields  a more clear picture of the proposed approximation. Moreover, the used approximation  avoids the analysis on the complex plane and is not always related to
the notion of the thermodynamic limit.
In the next sections we shall compare both approaches.

\vspace*{8pt}

\section{Steepest descent method and the calculation of the partition function}

By means of the well known identity on Gaussian integrals
\begin{equation}
e^{-\beta {\cal A}(\mathrm{u})^2}=\frac{1}{\sqrt{4\pi} i}\int_{-i\infty}^{+i\infty}d\lambda e^{-\beta^{1/2}{\cal A}(\mathrm{u})\lambda + \lambda^2/4},
\label{gi}
\end{equation}
Eq.~\eqref{nlt} may be presented 
in the form (the order
of  "$\lambda$" and "$\mathrm{u}$" integrations can be interchanged)

\begin{widetext}
\begin{eqnarray}
Z[H(\mathrm{u}; K_s)]=
\frac{1}{2\pi i}\left(\frac{\beta A K_s}{2}\right)^{1/2}\int_{-i\infty}^{i\infty}d\lambda \int{\it D}\{\mathrm{u}\}
\exp\Bigg\{-\beta\left[ {\cal T}(\mathrm{u})+\frac{\lambda}{\beta^{1/2}}{\cal A}(\mathrm{u})-\frac{\lambda^2}{4\beta}\right ]\Bigg \}.
\label{pfl}
\end{eqnarray}
\end{widetext}

In Eq.~\eqref{pfl} the shape fluctuation modes $(u^m_n)^2$ are decoupled due to Eq.~\eqref{gi}
so that the corresponding integrals become Gaussian.
Note that a similar procedure based on the integral representation Eq.~\eqref{gi} in the membrane theory has been used in \cite{Fap01,Bgp76,D97,Hi04}. However, working with complex functions is just one of the prices one has to pay to work with Gaussian integrals over fluctuation modes.

The partition function Eq. \eqref{pfl} may be rewritten in the  form:
\begin{eqnarray}
&&Z[H(\mathrm{u}; K_s)]=
\frac{1}{2\pi i}\left(\frac{\beta A K_s}{2}\right)^{1/2}\times \nonumber\\
&&\int_{-i\infty}^{i\infty}d\lambda \Bigg\{\int{\it D}\{\mathrm{u}\}
\exp [-\beta H(\mathrm{u};\lambda)]\Bigg\}.
\label{Zsm}
\end{eqnarray}

%\end{document}

The effective Hamiltonian in the exponential is given by:
%\begin{widetext|
\begin{eqnarray}
&&H(\mathrm{u};\lambda)=\nonumber\\
&&\frac{K_c}{2}\sum_{n=2}^{n_{max}}\sum_{m=-n}^n
(n-1)(n+2)\left[ n(n+1)+\right.
\left.\overline{\Sigma}(\lambda)\right](u^m_n)^2-
\frac{\lambda^2}{4\beta},\nonumber\\
\label{Ham}
\end{eqnarray}
%\end{widetext}
where
\begin{equation}
\overline{\Sigma}(\lambda):=\overline{\sigma}_0
+\overline{\sigma}_1(\beta)\lambda;\quad\overline{\sigma}_1(\beta)=\sqrt{\frac{ K_s}{2\beta A}}\frac {R_0^2}{K_c}.
\end{equation}
Now, if we look at the the partition function Eq.~\eqref{Zsm} %$Z[H(\mathrm{u}; K_s)]$,
we have the result
\begin{eqnarray}
&&Z[H
(\mathrm{u}; K_s)]=\nonumber\\
&&\frac{1}{2\pi i}\left(\frac{\beta A K_s}{2}\right)^{1/2}\int_{-i\infty}^{i\infty}d\lambda \exp\left\{ \frac{\lambda^2}{4}- \psi(\beta,\lambda) \right\},
\label{Zsm1}
\end{eqnarray}
where the following free energy has been introduced 
\begin{widetext}
\begin{eqnarray}
%\begin{split}
&&- \psi(\beta,\lambda)\equiv - \beta F[H(\mathrm{u};\lambda)] + \frac{\lambda^2}{4}=\nonumber\\
&&\ln\int{\it D}\{\mathrm{u}\}
\exp\Bigg \{-\beta \Bigg[\sum_{n=2}^{n_{max}}\sum_{m=-n}^n\frac{1}{2} K_c(n-1)(n+2)\Bigg(n(n+1)+\overline{\Sigma}(\lambda)\Bigg )\Bigg](u^m_n)^2\Bigg \}.\nonumber\\
\label{psi}
%\end{split}
\end{eqnarray}
\end{widetext}
%\end{document}
The exponential in Eq.~\eqref{psi} is a diagonal quadratic form in  the  real value  functions $(u_2^{-2}, u_2^{-1},\dots,
u_{n_{max}}^{n_{max}})$.   As a result the  multiple integral  over   $u^m_n$
splits into a product of one-dimensional  Gaussian integrals. These integrals over $u^m_n$   in Eq.\eqref{psi} can be performed easily. This statement is correct as long as
\begin{equation}
{\cal R}e\Bigg(n(n+1)+\overline{\Sigma}(\lambda)\Bigg) > 0,
\end{equation}
which implies ${\cal R}e (\overline{\Sigma}(\lambda)) >-6$.
Thus, we obtain the following expression
for  $\psi(\beta;\lambda)$: 
%\begin{widetext}
\begin{eqnarray}
&&\psi(\beta;\lambda)  
= \frac{N}{2}\ln\left(\frac{\beta K_c}{2\pi}\right)+\nonumber\\
 &&\sum_{n=2}^{n_{max}}
\frac{2n+1}{2} \ln\left\{(n-1)(n+2)[n(n+1)+ \overline{\Sigma}(\lambda)] \right\}
\mbox{}   \label{e62}.
\end{eqnarray}
%\end{widetext}

From Eqs.~\eqref{Zsm1},~\eqref{e62} and~\eqref{chv},
changing the variable 
%\end{document}
\begin{equation}
\frac{\lambda}{2 {\cal N}^{1/2}}=\xi, \quad {\cal N}\equiv \epsilon^{-1},
\label{chv}
\end{equation}

 we finally obtain:
\begin{widetext}
\begin{equation}
Z[H(\mathrm{u}; K_s)]=\frac{1}{2\pi i}\left(\frac{\beta A K_s}{2}\right)^{1/2}2{\cal N}^{1/2}\int_{-i\infty}^{i\infty}d\xi \exp\left\{{\cal N}[\xi^2- {\cal N}^{-1}\psi(\beta;2\xi{\cal N}^{1/2}) ]\right\}.
\label{Zs1a}
\end{equation}
\end{widetext}

%\end{document}
Let us briefly sketch the idea how one could compute the asymptotic expansion of the above integral, where ${\cal N} >>1$. If the term in the rectangular brackets in the exponential of Eq.~\eqref{Zs1a} has a finite limit $\Psi(\beta,\xi)$ provide ${\cal N} \to \infty$, the integral over $\xi$ in Eq.\eqref{Zs1a} can be carried out by the method of steepest descents. For this aim one needs to find the saddle point $\xi_0$ of the (holomorphic) function $ \Psi(\beta,\xi)$ and to deform the  path of integration in a way that does not affects its end points (as one may by Cauchy's theorem) so that it passes through the saddle point. Thus for the problem at hand one needs to deform the path of integration until the maximum of ${\cal R}e \Psi(\beta,\xi)$ along path of integration is also a stationary point of the $ {\cal I}m \Psi(\beta,\xi)$. One expects  the integral to be dominated by the saddle point $\xi_0$. 
Here, the following comments are in order. 

%\end{document}

First, one needs to compute $\Psi(\beta,\xi)$, i.e. the density free energy in the {\it thermodynamic limit}, which means $N\to\infty$ and $R_0 \to\infty$ while keeping $4\pi R_0^2/N $
constant. In this limit, the modes (l,m) are mapped onto wavevectors
${\bf q}$ contained in the plane $z=0$,  with the relation $q^2=[n(n+1)-2/R_0^2)]$, and we shall recover the planar model in the Fourier space (for details, see ref.\cite{GDM17}). The entire consideration  of the difference between
the free energy as given by the discrete sum of Eq.\eqref{e62} and as given  by the continuum expression $\Psi(\beta,\xi)$ indicates  that for a spherical membrane the finite-size effects are also important when are compared to the curvature ones \cite{MM94}. 

Second, precise justification of the method of steepest descent in our case needs some mathematical efforts.
It goes without saying that the integral in Eq.\eqref{Zs1a} can be replaced by the integrand: 
\begin{eqnarray}
&&Z[H(\mathrm{u}; K_s)]  \sim \nonumber\\
&& \Bigg[\exp{\cal N}\left\{\xi^2- {\cal N}^{-1}\psi(\beta;2\xi{\cal N}^{1/2}) \right\} + O({\cal N}^{-1})\Bigg],
\label{Zs1b}
\end{eqnarray} 
with $\xi$
obtained from the saddle point equation determining the extremum of the expression in the square  brackets:

\begin{equation}
\frac{d}{d\xi}\left\{\xi^2- {\cal N}^{-1}\psi(\beta;2\xi{\cal N}^{1/2}) \right\}=0,
\end{equation}
or equivalently
\begin{equation}
\xi=\frac{1}{2{\cal N}}\frac{d}{d\xi}\psi(\beta;2\xi{\cal N}^{1/2}),
\end{equation}
or equivalently in the explicit form
\begin{equation}
\lambda=\overline{\sigma}_1\sum_{n=2}^{n_{max}}\frac{2n+1}{n(n+1)+ \overline{\Sigma}(\lambda)}.
\label{eBT}
\end{equation} 
The free energy is now (up to an irrelevant constant)
\begin{equation}
F[H(\mathrm{u}; K_s)]=\frac{{\tilde\lambda}^2}{4}- \psi(\beta;{\tilde\lambda}),
\label{Fsp}
\end{equation}
where ${\tilde\lambda}$ is the solution of Eq.~\eqref{eBT}.

The study of the spectrum of thermally-excited shape fluctuations of bending and stretching deformations in vesicle membranes together,  is one of main topics of   this review. This could be performed in two alternative methods. First, as it is done so far, the membrane is considered to be compressible in the framework of a {\it relaxed version} of the conventional delta-function constraint. Then, one comes to a model effective Hamiltonian Eq.~\eqref{BTH} which is a result from Gaussian constrained membrane area (see,Eq.~\eqref{sc}). Second, in the next section, we shall consider a model where the area compressibility is taken into account  by additively added stretching energy term to the bending one. As a result, the self-consistent equation (see below) used to ensure a certain  membrane area constraint is found to be identical with the saddle point Eq.~\eqref{eBT}.
Furthermore we shall present an alternative method for calculating the partition function of the model Hamiltonian Eq.\eqref{BTH} that avoids the consideration in the complex plane \cite{BT19}.

%\end{document}

\section{The model Hamiltonian with an elastic contribution term.}

In this section we consider a vesicle whose membrane is made of a fixed number of constituent molecules N. Let us assume its {\it deformed/actual} surface area $S$  is governed by an elastic contribution to the model Hamiltonian. Following \cite{Bgp76}, (see also \cite{Fp03,Hi04}) the elastic contribution is proportional (with the coefficient of proportionality  the compressibility modulus $K_s$) to the square of the difference between the {\it actual} surface $S={N/\overline \rho}$ and {\it optimal} surface of the membrane $S_0=N/\rho_0$, where $ {\overline \rho}$ is the average value of the surface density of the constituent molecules in the deformed state, and 
$\rho_0$ is the average value of the surface density of the constituent molecules in the flat tension free (equilibrium)  state.  Optimal surface $S_0$ (called  also saturated or Schulman area \cite{Fp03}) 
is determined  by the intermolecular forces. Note that the effects of thermal
fluctuations  in the case of a nearly flat membrane using the Monge gauge have been studied  in \cite{Fp03}.

One can present the area functional of the actual (stretched)  membrane $S(\mathrm{v})$ in the
form:
\begin{equation}
S({\mathrm{v}})=4\pi R^2_0 +\Delta_ S({\mathrm{v}}), \label{r18}
\end{equation}
where the quantity
\begin{equation}
\Delta_ S(\mathrm{v})=  \frac{R^2_0}{2}
\Bigg[ \sum_{n=2}^{n_{max}}\sum_{m=-n}^n
(n-1)(n+2) (v_n^m)^2 \Bigg]  + O((v_n^m)^3). \label{e18}
\end{equation}
% $\Delta_ S({\mathrm{v}})$ 
is the
difference between the area of the vesicle's membrane and the area $4\pi R^2_0$
of a sphere with a reference volume equal to that of the vesicle, i.e. it is the dimensional excess area of the vesicle.
The symbol  ${\mathrm{v}}$ is used as a
shorthand for the real spherical harmonics amplitudes $(v_2^{-2}, v_2^{-1},\dots,
v_{n_{max}}^{n_{max}})$, appearing in
the expansion of the vesicle shape fluctuations from the equivalent volume sphere with radius $R_0$ (see Eq.~\eqref{e6}).
To point out the lack of the area constraint, we use the new notations for  the dynamical variables  $v^m_n$  instead  $u_n^m$ previously used in Eq.~\eqref{e6}
and  $S({\mathrm{u}})$ for the actual  area of the vesicle instead of $A({\mathrm{u}})$.

When the area functional 
$S(\mathrm{v})$ deviates (after stretching or compression) from the optimal area 
$S_0$ the membrane experiences a surface tension \cite{H73,L14}
\begin{equation}
\sigma(\mathrm{v}) = K_{s}\frac{S({\mathrm{v}})-S_{0}}{S_0} , \label{r17a}
\end{equation}
where $K_s$ is the area compressibility modulus.

The effective Hamiltonian
we  consider is presented as a sum of two terms:
\begin{equation}
H(\mathrm{v})=H_c(\mathrm{v})+H_s(\mathrm{v}), \label{1IB1}
\end{equation}
where (up to the order $(v_n^m)^2$)
\begin{equation}
H_c(\mathrm{v})=  \frac{1}{2}K_c \sum_{n=2}^{n_{max}}\sum_{m=-n}^n
(n-1)n(n+1)(n+2)(v_n^m)^2   \label{e17}
\end{equation}
for the   bending energy term, 
and
\begin{equation}
H_s(\mathrm{v})=\frac{S_0}{2K_s} [\sigma(\mathrm{v})]^2 \label{Hs1a}
\end{equation}
for the stretching energy term expressed via the membrane vesicle
tension
$\sigma(\mathrm{v})$, see refs.~\cite{H73,Bgp76,Fp03,Hi04,B10}.  

It will be instructive to clarify the relation of the considered here approach  with the approach of Milner and Safran and Seifert (see, Sec.4). The term in the rectangular brackets in Eq.~\eqref{Hs1a}
may be presented identically in the form
\begin{eqnarray}
&&[\sigma(\mathrm{v})]^2 =\nonumber\\ &&2\langle\sigma(\mathrm{v})\rangle_{H(\mathrm{v})}\sigma(\mathrm{v})-\langle\sigma(\mathrm{v})\rangle_{H(\mathrm{v})}^2+[\sigma(\mathrm{v})-\langle\sigma({\mathrm{v}})\rangle_{H(\mathrm{v})}]^2, \nonumber\\
\label{11495}
\end{eqnarray}
where the mean value is over the Hamiltonian Eq.~\eqref{1IB1}.
The last term
\begin{equation}
[\sigma(\mathrm{v})-\langle\sigma(\mathrm{v})\rangle_{H(\mathrm{v})}]^2
\label{msf}
\end{equation}
is the mean square fluctuations of the surface tension, which could be disregarded assuming that the fluctuations of the surface tension are small. 
%In addition, for the excess area $\Delta_S(\mathrm{v})$ as a %functional of $v_n^m$, let us make the assumption

Then, using Eqs.~\eqref{r18} and~\eqref{r17a}), one finds  
for the stretching energy functional:
\begin{equation}
H_s(\mathrm{v})\simeq H_{MF}(\mathrm{v}) = \langle\sigma(\mathrm{v})\rangle_{H(\mathrm{v})}S(\mathrm{v}) +\mathrm{const}.
\label{Ms1a}
\end{equation}
In fact, skipping the term Eq.~\eqref{msf} and keeping only the term linear with $(v^m_n)^2$ we come to a mean-field Hamiltonian $H_{MF}(\mathrm{v})$, similar to the one used in the Milner and Safran approach, see Eq.\eqref{HG}, in which the Lagrange multiplier 
$\sigma$ is replaced by $\langle\sigma(\mathrm{v})\rangle_{H(\mathrm{v})}$ \cite{B10,Fpab13}. If so, for the mean square amplitudes calculated with $H_{MF}(\mathrm{v})$ we get
\begin{equation}
\left\langle|v^m_n|^2\right\rangle_{H_{MF}(\mathrm{v})}=\frac{8\pi \gamma}{(n-1)(n+2)[n(n+1)+\overline {\langle\sigma(\mathrm{v})\rangle}_{H(\mathrm{v})}]}.
\label{MSfu}
\end{equation}

Here, the following comments are in order. Indeed, using parameter $\gamma$ (respectively  $K_c$) and dimensionless $\overline {\langle\sigma(\mathrm{v})\rangle}_{H(\mathrm{v})}$  as fitting  parameters, one can infer information  about them from the flickering analysis, but the information  about $K_s$ (i.e. 
the role of the stretching effects) on the thermal fluctuation of the membrane remains  hidden.
The main problem is that we do not calculate the mean of $\sigma(\mathrm{v})$ with the Hamiltonian $H(\mathrm{v})$   in the denominator of the rhs of Eq.~\eqref{MSfu} (it is possible self-consistently only in the mean-field approximation) and  we can't estimate the error we made where skipping the last term in Eq.~\eqref{11495}. In what follows we shall try to
shed light on these problems.

Since the Hamiltonian, Eq.~\eqref{1IB1} (with Eqs.~\eqref{r18} and~\eqref{Hs1a}) is {\it exactly the same} as Eq.~\eqref{BTH} with $A$ replaced by $S_0$ we may use for convenience the equivalent
(up to an irrelevant constant) presentation
\begin{equation}
H(\mathrm{v})={\cal T}({\mathrm{v}})+[{\cal A}(\mathrm{v})]^2, \label{z148}
\end{equation}
where ${\cal T}(\mathrm{v})$ and ${\cal A}(\mathrm{v})$ are defined by 
Eqs.~\eqref{e140} and~\eqref{e16}, respectively.
The last term  due to its nonlinearity with respect to the squares
of the amplitudes $v_n^m$  causes computational problems which as it was shown in the previous section can be solved using the steepest descent method. Below,  to overcome  this
obstacle we shall follow a different way. We shall linearize the Hamiltonian \eqref{z148}
using "approximating Hamiltonian method" (about this method, see, e.g. Ch.2 in \cite{B00} and \cite{B81,B84,Bj13}) based on the Bogoliubov  variational inequalities. This approach might be an advantage over the former,
because it  avoids  computations in the complex plane (and in some sense the direct implementation of the thermodynamic limit procedure).

%\end{document}
\section{The approximating Hamiltonian and the calculation of the partition function.}

The idea of the proposed approximation is to replace the non-solvable initial Hamiltonian $ H(\mathrm{v})$ with a more simple linearized  Hamiltonian $H_{app}(\mathrm{v},X)$ depending on a variational parameter $X$. The resulting Hamiltonian is called  "approximating Hamiltonian" if under a proper   choice of its free parameter $X$ it can be proved asymptotically closer to the initial   one, in the sense that both Hamiltonians  generate the same thermodynamic behaviour. Thus the problem of interest is reduced to a simpler one, which allows to obtain its thermodynamic functions in an analytical form. 
Below we shall realize this program following our studies in ref. \cite{BT19}.

The second term in Eq.~\eqref{e141} may be presented in the form:
\begin{equation}
[{\cal A}(\mathrm{v})]^2 = 2X{\cal A}(\mathrm{v})-X^2+[{\cal A}(\mathrm{v})-X]^2, \label{e142}
\end{equation}
where  $X$ is an arbitrary real parameter. We define the linearized Hamiltonian
$H_{app}(\mathrm{v},X)$ as:
\begin{equation}
H_{app}(\mathrm{v},X)= {\cal T}(\mathrm{v}) +2X{\cal A}(\mathrm{v})-X^2. \label{e44}
\end{equation}
The last equation is obtained from Eq.~\eqref{e141} by removing the term $[{\cal A}(\mathrm{v})-X]^2$ from the right-hand-side of Eq.~\eqref{e142}.  The problem is to prove that  the skipped term is in some sense small. Then the defined in this way Hamiltonian $H_{app}(\mathrm{v},X)$ is linear with respect to $(u_n^m)^2$ and the corresponding partition function 
\begin{eqnarray}
&&Z[H_{app}(\mathrm{v},X)]:=\exp\{-\beta  F[H_{app}
(\mathrm{v},X)]\}=\nonumber\\
&&\int{\it D}\{\mathrm{v}\}
\exp\Bigg\{-\beta\left[ {\cal T}(\mathrm{v})+2X{\cal A}(\mathrm{v})-X^2\right ]\Bigg \}.
\label{pf2}
\end{eqnarray}
is a trivial Gaussian integral. 

A simple comparison of the terms in the curly brackets in 
Eqs.~\eqref{pfl} and \eqref{pf2} shows that they define the same 
(up to an irrelevant factor) partition function
provided $\lambda=2\beta^{1/2}X$. 
In this sense, the use of the soft-area constraint Eq.~\eqref{sc} is equivalent to the inclusion of the stretching energy term Eq.~\eqref{Hs1a} in the Hamiltonian.
However,  the crucial difference is that the arbitrary complex parameter $\lambda$ and the real parameter $X$ should be fixed under different rules. 

Anyway, we may perform the integrations in Eq.~\eqref{pf2} directly by using the result Eq.~\eqref{psi}, which for the free energy gives:
\begin{widetext}
\begin{eqnarray}
F[H_{app}(\mathrm{v},X)]= k_BT \sum_{n=2}^{n_{max}}
\frac{2n+1}{2} \ln\left\{(n-1)(n+2)[n(n+1)+ \overline{\Sigma}_{app}(X)] \right\}
\mbox{} -
&&X^2 + \frac{N}{2}\ln\left(\frac{\beta K_c}{2\pi}\right) \label{2e6},
\end{eqnarray}
\end{widetext}
where 
\begin{equation}
\overline{\Sigma}_{app}(X):=\overline{\sigma}_0
+\sigma_1X,
\label{d83}
\end{equation}
with
\begin{equation}
\overline{\sigma}_0=K_s\frac{R^2_0}{K_c}\left(\frac{4\pi R^2_0}{S_0}-1\right),
\quad \sigma_1=\sqrt{\frac{2 K_s}{S_0}}\frac {R_0^2}{K_c}.
\label{e1s6}
\end{equation}

Our goal is  to develop an approach which allows to control the approximation~\eqref{2e6}  via the parameters of the considered model. In what follows we shall prove that 
%instead of $F[H_{app}(\mathrm{v},K_s)] $
it is possible to work with $F[H_{app}(\mathrm{v},X)]$ provided the parameter $X$ is fixed in an appropriate way.

%\end{document}
\section{The approximating Hamiltonian and Bogoliubov variational inequalities}

In their most convenient form the Bogoliubov variational inequalities  are given by
\begin{widetext}
\begin{eqnarray}
\langle  H-H_{app}(X) \rangle_{H}  
\le  F[H]-F[H_{app}(X)] \le  \langle H-H_{app}(X) \rangle_{H_{app}(X)},
\label{e46}
\end{eqnarray}
\end{widetext}
where   $F[H]$ is the free energy  of a valid Hamiltonian $H$ and 
$F[H_{app}(X)]$ is the free energy
of a presumably simpler Hamiltonian $H_{app} (X)$, depending on a variational parameter $X$. 
The variational parameter $X$
must be determined from the condition of the best approximation
of $F[H]$. As we said earlier in the case of the best approximation $H_{app} (X)$ is called Approximating Hamiltonian.

It is worth noting that, although the two-side estimate \eqref{e46}   is almost an evident consequence of the convexity of the free energy, its    proof on a rigorous level needs significant mathematical efforts, for details  see Sec. 3.4, and for historical remarks Sec.3.5 in ref. \cite{Z20}. 

The recipe for the determination of the Approximating Hamiltonian in combination with the Bogoliubov variational inequalities is the essence of the so-called approximating Hamiltonian method (AHM).
The advantage of the method is that in many cases it is possible to estimate the correlator in the left and right side of the inequalities  \eqref{e46}. This gives a further insight into the common approximation, usually based on physical intuition, and leads to new
result as well.
There exists an extensive literature about the AHM, see, e.g.  Ch.2 in \cite{B00},\cite{B81,B84,Bj13} for a list of different applications in the theory of critical phenomena and condensed matter physics. 

The following comment is in order here. The second of the inequalities Eq.~
(\ref{e46}) is known as Bogoliubov variational upper bound of the exact free energy~\cite{T67}, (see also Ch.2 in \cite{S03}). Using only this part of the inequalities \eqref{e46} the best approximation from above is obtained, if the variational parameter $X$ {\it minimizes} the variational free energy $F_{var}(X)$, defined as follows:
\begin{equation}
F_{var}(X):= F[H_{app}(X)] +  \langle H-H_{app}(X) \rangle_{H_{app}(X)}.
\label{c12a}
\end{equation}
This allows to obtain approximation  from above 
\begin{equation}
F[H]\le \min_X F_{var}(X)
\end{equation}
(although some times quite crudely)  for the exact free energy of the studied  physical system. The approach based on (\ref{c12a}) is effectively used in order to obtain closed-form expression for the spectra of the thermal fluctuation of spherical vesicles incorporating nonlinear curvature elasticity terms \cite{AS16}.

We shall use another quite different approach which allows to estimate the used approximation. If the lhs of  (\ref{e46}) is positive definite, the best approximation of $f[H]$ from below is obtained
{\it maximizing}  $f[H_{app}(X)]$ with respect to $X$. In this case one can estimate the approximation through the estimation of the  thermodynamic mean value in the rhs of  (\ref{e46}). The use  of  inequalities (\ref{e46}) in the  statistical mechanics of a lipid vesicle has been announced  in \cite{BT14}.

In the inequalities \eqref{e46} we choose
\begin{equation}
H:=H(\mathrm{v};K_s)={\cal T}(\mathrm{v})+[{\cal A}(\mathrm{v})]^2, \label{e1z}
\end{equation}
and its linearized version as:
\begin{equation}
H_{app}:=H_{app}(\mathrm{v},X)= {\cal T}(\mathrm{v}) +2X{\cal A}(\mathrm{v})-X^2. \label{e2}
\end{equation}

Since the thermal average of a nonnegative quantity is nonnegative, it follows that:
\begin{eqnarray}
\langle H(\mathrm{v};K_s)- H_{app}(\mathrm{v},X) \rangle_{H(\mathrm{v})} 
=\langle [{\cal A}(\mathrm{v})-X]^2 
\rangle_{H(\mathrm{v})} \ge 0.\nonumber\\
 \label{e147}
\end{eqnarray}
Then Eqs.~\eqref{e46} and \eqref{e147} imply that for each $X$:
\begin{equation}
0 \le F[H(\mathrm{v};K_s)] - F[H_{app}(\mathrm{v},X)] \le \langle [{\cal A}(\mathrm{v})-X]^2
\rangle_{H_{app}(\mathrm{v},X)}. \label{e148}
\end{equation}
Now, we can determine $X$ from the condition of the best approximation
\begin{equation}
F[H_{app}(\mathrm{v},\widetilde{X})]=\max_X F[H_{app}(\mathrm{v},X)]. \label{e149}
\end{equation}
Hereafter, in order to be unambiguous  we shall use  notations linearized and
approximating Hamiltonian
for $H_{app}(\mathrm{v},X)$ and
$H_{app}(\mathrm{v},\widetilde{X})$, respectively.

Since $F[H_{app}(\mathrm{v},X)]$ is a differentiable function of $X$,   thus $\widetilde{X}$ is defined as  solution of the equation:
\begin{equation}
%\left.
\frac{\partial F[H_{app}(\mathrm{v},X)]}{\partial X}=0.
\label{eq51}
\end{equation}
Differentiating 
\begin{eqnarray}
&&F[H_{app}(\mathrm{v},X)]  = -X^2-\nonumber\\
&& \beta^{-1} 
\ln{\left\{\int D\{\mathrm{v}\}\exp\{ -\beta\left[{\cal T}(\mathrm{v})+2X{\cal A}(\mathrm{v})\right]\}\right\}} , \label{fr1)}
\end{eqnarray}
we obtain:
\begin{eqnarray}
\frac{\partial F[H_{app}(\mathrm{v},X)]}{\partial X} 
=2[\langle {\cal A}(\mathrm{v}) \rangle_{H_{app}(\mathrm{v},X)} -X]=0.\nonumber\\
\label{e52}
\end{eqnarray}
This is a typical self-consistent equation for the variational parameter "$X$".

Differentiating the lhs of Eq.~\eqref{e52} ones more, we obtain
\begin{eqnarray}
&&\frac{\partial^2 F[H_{app}(\mathrm{v},X)]}{\partial^2 X} 
=\nonumber\\
&&-2\{\beta \left \langle \left[\langle {\cal A}(\mathrm{v})\rangle_{H_{app}({\cal A},X)} - {\cal A}(\mathrm{v}) \right]^2\right\rangle_{H_{app}(\mathrm{v},X)} + 1\} < 0.\nonumber\\
\label{v52}
\end{eqnarray}
Consequently, $F[H_{app}(\mathrm{v},\widetilde{X})]$ is a convex function of $X$ and thus Eq.\eqref{e52} has only one solution, namely $\widetilde{X}$. 
From Eq.~\eqref{e148} it follows that:
\begin{equation}
F[H_{app}(\mathrm{v},\widetilde{X})]\le F[H(\mathrm{v})].
\end{equation}
Thus we show that the free energy $F[H_{app}(\mathrm{v},\widetilde{X})]$ of the ensemble of
not
"interacting amplitudes $u^m_n$" described by the Hamiltonian \eqref{ex} is the best approximation from below of the free energy,
corresponding to the model Hamiltonian $H(\mathrm{v};K_s)$. It is important to note that the correlator in the r.h.s. of the inequalities
\eqref{e148} can be calculated and thus to estimate the approximation of $F[H(\mathrm{v};K_s)]$ through $F[H_{app}(\mathrm{v},\widetilde{X})]$ .

%\end{document}

\section{Analysis of the self-consistent equation} 

The mean squares amplitudes  $\langle (u_n^m)^2\rangle_{H_{app}(\mathrm{v},X)}$, calculated
by the linearized Hamiltonian $H_{app}(\mathrm{v},X)$ (compare 
with Eq.~\eqref {Ham}), 
\begin{eqnarray}
&&H_{app}(\mathrm{v};X)= - X^2+
\frac{1}{2}\sum_{n=2}^{n_{max}}\sum_{m=-n}^nK_c
(n-1)(n+2)\times\nonumber\\
&&\left\{ n(n+1)+\overline{\sigma}_0
+X\sqrt{\frac{2 K_s}{\beta A}}\frac {R_0^2}{K_c
}\right\}|
(v^m_n)^2.
\label{ex}
\end{eqnarray} 
are:
\begin{eqnarray}
\langle (v_n^m)^2\rangle_{H_{app}(\mathrm{v},X)}
=\frac{8\pi \gamma}{(n-1)(n+2)[n(n+1)+\overline{\Sigma}_{app}(X)]}.\nonumber\\
\label{e670}
\end{eqnarray}

As it can be seen after comparison with Eq.~\eqref{MSf}, the
result for $\langle (v_n^m)^2\rangle_{H_{app}(\mathrm{v},X)}$ formally reproduces the result of Milner and Safran. However, a significant difference  takes place. In their theory  $\overline{\sigma}_{MS}$ is introduced  as a
Lagrange multiplier, while here $\overline{\Sigma}_{app}(X)$ with $X=\widetilde{X}$ is obtained self-consistently from $F[H_{app}(\mathrm{v},\widetilde{X})]$.

After differentiating Eq.~\eqref{2e6} with respect to $X$ we obtain Eq.~\eqref{eq51} in an explicit form (compare with Eq.~\eqref{eBT})
\begin{equation}
X=\frac{kT\sigma_1}{4}\sum_{n=2}^{n_{max}} \frac{2n+1}{n(n+1)+
\overline{\sigma}_{0}+\sigma_{1}X}. \label{e68}
\end{equation}
This equation is the analogue in our case  of the equation, obtained in \cite{Bgp76} and \cite{D97}, for the renormalized surface tension of almost planar membranes.
Using the definition Eq.~\eqref{d83} we shall rewrite Eq.~\eqref{e68} 
in a  more convenient form:
\begin{equation}
\overline{\Sigma}_{app} =
\overline{\sigma}_0+
\overline{C}\sum_{n=2}^{n_{max}} \frac{2n+1}{n(n+1)+\overline{\Sigma}_{app}},
\label{e68c}
\end{equation}
where  the shorthands
\begin{equation}
\overline{\Sigma}_{app} = \overline{\Sigma}_{app}(\widetilde{X}).
\label{53}
\end{equation}
and
\begin{equation}
\overline{C}=
\frac{1}{2\beta}\frac{K_s}{S_0}\frac{R^4_0}{(K_c)^2}\approx \gamma K_s\frac{R^2_0}{K_c}.
\label{C}
\end{equation}
are used. In the numerical computations sometimes it is reasonable to use the last term in Eq.~\eqref{C} obtained after the approximation $R^2_0/S_0 \approx 1/4\pi$.  

In the general case Eq.(\ref{e68c}) can be solved only numerically.
Note that it is possible first to obtain $\overline{\Sigma}_{app}(K_c,K_s)$ from Eq.~\eqref{e68c}, indeed numerically, and after that to calculate from Eq.~\eqref{e670} the dependence of $\langle(v_n^m)^2\rangle_{H_{app}(\mathrm{v},\widetilde{X})}$ on $K_c$,  $ K_s$.

Further on for the sake of simplicity for the solution of Eq.~\eqref{e68c} we shall use the notation $\overline{\Sigma}_{app}(\overline{C},\overline{\sigma}_0,N\approx n^2_{max})=\overline{\Sigma}_{app}$.
Eq.~\eqref{e68c} shows that $\overline{\Sigma}_{app}$ depends on  $K_s$ and $K_c$ and geometrical parameters $R_0$ and $S_0$ only in the combinations $\overline{C}$ and $~\overline{\sigma}_0$.
Under the condition
\begin{equation}
-\frac{\overline{\sigma}_0}{ \overline{C}}\equiv \frac{1}{\gamma}\frac{S_0-4\pi R_0^2}{4\pi R_0^2}\le\sum_{n=2}^{n_{max}} \frac{2n+1}{n(n+1)},
\label{con1}
\end{equation}
Eq.~\eqref{e68c} has only one solution which belongs to the interval $[0,\infty)$.
If the opposite inequality 
takes place, $\overline{\Sigma}_{app}$ belongs to the interval $(-6,0]$. Let us present some numerical results for $\overline{\Sigma}_{app}$. For example, if $n_{max}=3.10^4$ , the quantity in the rhs of the inequality \eqref{con1} is $19.27... $. In this case, under the condition $\overline{C}=10^5$, if $-\frac{\overline{\sigma}_0}{ \overline{C}}=19.27...$ the solution is
$\overline{\Sigma}_{app}=0$. If $-\frac{\overline{\sigma}_0}{ \overline{C}}=16.50...$ the solution is $\overline{\Sigma}_{app}=57,72$, and if $-\frac{\overline{\sigma}_0}{ \overline{C}}=19.50...$ the solution $\overline{\Sigma}_{app}=57,72=-0,82$. Note that  the ratio $\overline{\sigma}_{0}/\overline{C}$ does not depend on $K_s$ and as follows from inequality \eqref{con1} the sign of the solution $\overline{\Sigma}_{app}$ does not depend on $K_s$.

For $\overline{\Sigma}_{app}>>1$, Eq.\eqref{e68c} can be solved analytically  in terms of the Lambert  function (see Eq.~\eqref{Lsol} in the Appendix). 
In this case  two different regimes  have to be distinguished (see the Appendix): 
%\end{document}
a)	
\begin{equation}
\overline{\Sigma}_{app}=\overline{C}\ln  \left[\frac{N\exp(\overline{\sigma}_{0}/\overline{C})}{\overline{C}}\right]
,\qquad \frac{N\exp(\overline{\sigma}_{0}/\overline{C})}{\overline{C}}>>1,
\label{007}
\end{equation}
or

b)
\begin{equation}
\overline{\Sigma}_{app}= N\exp(\overline{\sigma}_{0}/\overline{C})
,\qquad \frac{N\exp(\overline{\sigma}_{0}/\overline{C})}{\overline{C}}<<1,
\label{Lso18}
\end{equation}
The former case has to be attributed to
finite  $K_s$, while the latter to the limit case $K_s \to \infty$.

Now it is convenient to include an auxiliary effective tension $\overline{\Sigma}_{MS}$  related to  a {\it reference vesicle with fixed area and volume}  $A=S_0$, $V=\frac{4\pi}{3}R^3_0$ and with an excess area $\Delta_{MS} $  defined in Eq.~\eqref{b101} (i.~e.~with the same values as those of the vesicle considered in Section 4 which is the reason to use the same notation $\overline{\Sigma}_{MS}$). 	
Thus, the definitions Eqs.~\eqref{a1},~\eqref{g1},~\eqref{e16} and~\eqref{C} employ the identity 

\begin{equation}
-\frac{\Delta_{MS}}{\gamma}=\frac{\overline{\sigma}_0}{\overline{C}},\quad \overline{\sigma}_0<0.
\label{rconst1}
\end{equation} 
Now it is possible
to insert the value of $\overline{\Sigma}_{MS}$ from Eq.\eqref{sol02} in Eqs.(\ref{007}) and (\ref{Lso18}). One gets:

%a')	
\begin{equation}
\overline{\Sigma}_{app}=\overline{C}\ln  \left(\frac{\overline{\Sigma}_{MS}}{\overline{C}}\right),
\qquad \frac{N\exp(\overline{\sigma}_{0}/\overline{C})}{\overline{C}}>>1
\label{Lsol9a},
\end{equation}

or
%b')
\begin{equation}
\overline{\Sigma}_{app}=\overline{\Sigma}_{MS}
,\qquad \frac{N\exp(\overline{\sigma}_{0}/\overline{C})}{\overline{C}}<<1.
\label{Lso18b}
\end{equation}

Thus, one obtains $1<< \overline{\Sigma}_{app}\le\overline{\Sigma}_{MS}$. 
Eqs. (\ref{Lsol9a}) and (\ref{Lso18b})  allow to keep track of the two effective 
tensions, $ \overline{\Sigma}_{app}$ and $\overline{\Sigma}_{MS}$, under the condition $\exp(-\Delta_{MS}/\gamma)<<1$, which validates the result Eq.~\eqref{sol02} .

%\end{document}
\section{The surface tension}

It is worth noting that in the presented approach $\overline{\Sigma}_{app}$  has a natural physical interpretation: Eq.~\eqref{e68c} (recall the relations Eqs.~\eqref{r18}, ~\eqref{e18} and \eqref{e670}) implies:

\begin{eqnarray}
\overline{\Sigma}_{app}=\frac{R^2_0}{K_c} 
K_s\frac{\langle
S(\mathrm{v})\rangle_{H_{app}(\mathrm{v},\widetilde{X})}-S_0}{S_0}
=\frac{R^2_0}{K_c}
\langle \sigma(\mathrm{v}) \rangle_{H_{app}(\mathrm{v},\widetilde{X})}, \label{ks1}\nonumber\\
\end{eqnarray}
where $\sigma(\mathrm{v})$ is the (not normalized) tension of the membrane (see
Eq.~\eqref{r17}).
In the Appendix B it is shown that if the free energy of the initial and the approximating Hamiltonian are thermodynamically equivalent, 
then we have for the true (calculated with $H(\mathrm{v})$) tension (see the definition Eq.~\eqref{r17a}):
\begin{equation}
\overline{\Sigma}_{app} \to \left(\frac{R^2_0}{K_c}\right)\langle \sigma(\mathrm{v})\ \rangle_{H(\mathrm{v})} . 
\end{equation}

Let us replace 
$\overline{\Sigma}_{app}$ 
in favor of the other meaningful quantity- 
dimensionless	excess area $\Delta(\overline{\Sigma}_{app})$:
\begin{equation}
\Delta(\overline{\Sigma}_{app})\equiv \frac{\langle
S(\mathrm{v})\rangle_{H_{app}(\mathrm{v},\widetilde{X})}-4\pi R^2_0}{4\pi R^2_0}
\end{equation}
From Eqs.~\eqref{r18},\eqref{e18} and \eqref{e670} (with $X=\widetilde{X}$) it follows that:
\begin{equation}
\Delta(\overline{\Sigma}_{app})   =
\gamma\sum_{n=2}^{n_{max}}
\frac{2n+1}{n(n+1)+\overline{\Sigma}_{app}}.
\label{e78}
\end{equation}

Using the Eq.~\eqref{sol01} obtained in the Appendix A the rhs of the Eq.(\ref{e78}) may be transformed in a more simple form. Thus one gets:
\begin{equation}
\Delta(\overline{\Sigma}_{app})\approx\gamma \ln\left(\frac{N}{\overline{\Sigma}_{app}}
\right),\qquad \frac{N}{\overline{\Sigma}_{app}}>>1.
\label{so222}
\end{equation}	

Now, it is possible to compare in an explicit form $\Delta(\overline{\Sigma}_{app})$ with the excess area obtained  within the approach of Milner and Safran $\Delta(\overline{\Sigma}_{MS})$, see Eq.~\eqref{so202}. One gets  

\begin{equation}
\Delta(\overline{\Sigma}_{app})-\Delta(\overline{\Sigma}_{MS})\approx\gamma \ln\left(\frac{\overline{\Sigma}_{MS}}{\overline{\Sigma}_{app}}
\right)\ge 0
\label{so232}
\end{equation}	
indeed, under the conditions:
\begin{equation}
\frac{\overline{\Sigma}_{app}}{N}\le\frac{\overline{\Sigma}_{MS}}{N}<<1, \qquad 1<< \overline{\Sigma}_{app}\le\overline{\Sigma}_{MS}.
\label{msap}
\end{equation} 	

Moreover, provided the inequality $\overline{\Sigma}_{app}\le\overline{\Sigma}_{MS}$ takes place, by comparing  the Eqs.~\eqref{b101} and ~\eqref{e78} one may conclude that
always $\Delta(\overline{\Sigma}_{app}) \ge \Delta(\overline{\Sigma}_{MS})$. Thus, as follows from Eqs.~\eqref{b101} and ~\eqref{e78}, a larger $ K_s$ value will result in
a smaller excess area $\Delta(\overline{\Sigma}_{app})$.

%\end{document}
\section{The relation between $\overline{\Sigma}_{app}$ and $K_s$}

From Eq.~\eqref{b101} with $A=S_0$ one can see that the effective tension $\overline{\Sigma}_{MS}$ obeys the equation:
\begin{equation}
0 =
\overline{\sigma}_0+
\overline{C}\sum_{n=2}^{n_{max}} \frac{2n+1}{n(n+1)+\overline{\Sigma}_{MS}}.
\label{e68d}
\end{equation}
Now, with the help of Eq.~\eqref{e68d} the Eq.~\eqref{e68c} may be presented in the form:
\begin{eqnarray}
&&\overline{\Sigma}_{app}=
%\frac{R^4}{2K_c}\frac{K_s}{S_0}\frac{kT}{K_c}
\overline{C}(\overline{\Sigma}_{MS}-\overline{\Sigma}_{app})\times\nonumber\\
&&\sum_{n=2}^{n_{max}}
\frac{(2n+1)}{[n(n+1)+\overline{\Sigma}_{app}]
[n(n+1)+\overline{\Sigma}_{MS}]}.
\label{bt2} 
\end{eqnarray}
Eq.~\eqref{bt2} shows the dependence of   $\overline{\Sigma}_{app}$ as a function of $K_c$, $K_s$ and $\overline{\Sigma}_{MS}$ (respectively  $S_0$) at fixed $R_0$ and $T$.
For $\overline{\Sigma}_{app}>>1$, this dependence may be obtained in   terms of the Lambert  function (see Eq.~\eqref{Lwol} in the Appendix A).

The simple relation between $\overline{\Sigma}_{app}$ and $K_s$ prompts Eq.~\eqref{bt2}
to be inverted to yield  $K_s$  as
a function of $\overline{\Sigma}_{app}$:
\begin{eqnarray}
\frac{R^2_0}{K_c}K_s=
{\cal F}_s(\overline{\Sigma}_{MS},\overline{\Sigma}_{app}),
\label{Ks}
\end{eqnarray}
where
\begin{eqnarray}
&&{\cal F}_s(\overline{\Sigma}_{MS},\overline{\Sigma}_{app})=\gamma
\frac{\overline{\Sigma}_{app}}{\overline{\Sigma}_{MS}-
\overline{\Sigma}_{app}}\times\nonumber\\  
&&\left \{\sum_{n=2}^{n_{max}}
\frac{(2n+1)}{[n(n+1)+\overline{\Sigma}_{MS}]
[n(n+1)+\overline{\Sigma}_{app}]}\right \}^{-1}.
\label{e68e}
\end{eqnarray}

Note that by definition the stretching modulus $K_s$ is positive. Since negative values of $\overline{\Sigma}_{MS}$ are allowed \cite{S95} the following two possibilities  depending on the sign   of  $\overline{\Sigma}_{MS}$ in Eq.\eqref{e68e}  are relevant: 

a) $\overline{\Sigma}_{MS} <\overline{\Sigma}_{app}<0,$ when  $-6<\overline{\Sigma}_{MS} <0 $,

b) 
$\overline{\Sigma}_{MS} >\overline{\Sigma}_{app}>0 $, when  $\overline{\Sigma}_{MS} >0 $.

The obtained result  determines   the values of  $\overline{\Sigma}_{app}(K_s)$ when $K_s\rightarrow 0$ and $K_s \rightarrow \infty$ (when all the other model parameters are fixed). 

For the first limit the result is:
\begin{equation}
\lim_{K_s\rightarrow 0}\overline{\Sigma}_{app}(K_s)=0. \label{bt3}
\end{equation}
Trivially, the analogous case in the Milner and Safran approach is the case when the Lagrange multiplier $\sigma=0$. 

Since $\overline{\Sigma}_{MS}$ does not depend on $K_s$, from the rhs of Eq.~\eqref{e68e} it follows that when $\overline{\Sigma}_{app} \rightarrow \overline{\Sigma}_{MS}$, one gets $K_s \rightarrow \infty$. 
The second limit is exactly the  tension  of the reference incompressible membrane $\overline{\Sigma}_{MS}$ (see, also Eq.\eqref {Lwol}
):
\begin{equation}
\lim_{K_s\rightarrow \infty}\overline{\Sigma}_{app}(K_s)= \overline{\Sigma}_{MS}. \label{bt4}
\end{equation}
The theory of Milner and Safran  is adequate within  regimes in which $K_s$ is not  relevant.  Eqs.~\eqref{bt3} and ~\eqref{bt4} suggest that  it is a limiting  case of the presented theory. 
We showed that the self-consistent equation allows to obtain the stretching elasticity modulus $K_s$ via experimentally accessible quantities.
% Indeed, from Eqs. \eqref{e670}, \eqref{e68d0}, and \eqref{bt10}

%\end{document}

\section {The fitting function}

In this section we shall discuss  the connection of the above theory with experimental studies of vesicle fluctuations in the context of  of flicker-noise measurement as well. 

First of all, it is instructive   to estimate the constants in $\gamma,\overline{\sigma}_0$ and $\overline{C}$. We shall  use the following typical numerical values of the quantities~\cite{Mel97} entering in our model Hamiltonian:\\
$K_s\sim100$~erg/cm$^2$;
$\quad K_c\sim10^{-12}$erg;
$\quad R_0\sim10^{-3}$cm;
$\quad S_0\sim 4\pi R_0^2\sim 1.256\times 10^{-5}$cm$^2$;\\
$\overline{\sigma}_1\equiv (R_0^2/K_c)\sigma_1=4\times 10^9$erg$^{-0.5}$;
$\quad k_BT \sim 4.10^{-14}$erg.\\ 
We accept for the estimation of membrane stretching the typical value $\sigma_0 \sim$ 1~erg/cm$^2$ and
obtain the following values:\\
$\Delta_{MS} \sim  10^{-2}$ ;
$\quad \gamma  \sim 10^{-3}$ ;
$\quad|\overline{\sigma}_0| \sim 10^6$ ;
$\quad \overline{C}\sim 10^{5}$. \\
Evidently, the above constants obey the relation 
\begin{equation}
\frac{\Delta_{MS}}{\gamma}=\frac{|\overline{\sigma}_0|}{\overline{C}}\sim 10.
\label{rconst}
\end{equation} 
Thus, the inequality Eq.~\eqref{con1} holds and Eq. (\ref{e68c}) has a positive solution for $\overline{\Sigma}_{app}$. Let us recall that two regimes resulting from the inequality between $\overline{C} $ and $N$ are  possible: i) given by Eq.~(\ref{007}) (or alternatively Eq.~\eqref{Lsol9a}), or ii)	given by Eq.~(\ref{Lso18}) (or alternatively Eq.\eqref{Lso18b}).
In the above statement we accept also that the intermolecular distance $\lambda$ is of the order of 10 \AA
~and then $n_{max}\sim 3.10^{4}$ and $N\sim 10^9$.

In the general case $\overline{\Sigma}_{app}$ depends on $K_c$, $K_s$, $R_0$, $S_0$ and $N$. If  $R_0$ and $S_0$ can be calculated or measured by independent methods and are not correlated with $K_c$ and $K_s$, then, by fitting $\langle(v_n^m)^2\rangle_{H_{app}(\mathrm{v},\widetilde{X})}$ with $K_c$ and $K_s$, we can determine them   from the analysis of the thermal fluctuations of the vesicle shape.
However, to perform the needed fitting procedure is a highly non trivial task, since the fitting parameters are contained in an implicit form in  $\overline{\Sigma}_{app}$. 

%\end{document}

A straightforward way to obtain the fitting function in an explicit form  is to use the approximative solution $\overline{\Sigma}_{app}$ given by Eq.\eqref{Lwol} in the Appendix A in the rhs of Eq.\eqref{e670}. 
%А visible simplification takes place if one use the approximating %expression given in Eq.\eqref{Lsol9a} instead. 
In addition, in terms of the approximative result Eq.\eqref{Lsol9a}, one gets
%\end{document}
\begin{eqnarray}
&&\langle (v_n^m)^2\rangle_{H_{app}(\mathrm{v},X)}\approx\nonumber\\
&&\frac{8\pi\gamma}{(n-1)(n+2)\left\{n(n+1)+\gamma\overline{K}_{s}[\ln\overline{\Sigma}_{MS}-\ln(\gamma\overline{K}_{s})]\right\}},\nonumber\\
\label{e670n}
\end{eqnarray}
where $\overline{K}_{s}=\frac{R^2}{K_c}K_s$ is the dimensionless area compressibility  modulus. Recall that (see Eq.~\eqref{Lsol9a}), the above equation becomes valid provided that the condition
\begin{equation}
\frac{\overline{\Sigma}_{MS}}{\gamma\overline{K}_{s}}>>1
\label{nfs}
\end{equation}
takes place. As follows from Eq.~\eqref{Lso18b},  the opposite strong inequality provides  the case considered by Milner and Safran:
\begin{eqnarray}
\langle (v_n^m)^2\rangle_{H_{app}(\mathrm{v},X)}=
\frac{8\pi\gamma}{(n-1)(n+2)\left\{n(n+1)+\overline{\Sigma}_{MS}\right\}},\nonumber\\
\label{e670m}
\end{eqnarray}
Eqs.~\eqref{e670n}, \eqref{nfs} and \eqref{e670m} for the mean-square amplitudes could be used to determine experimentally $K_s$, $\gamma$ (respectively $K_c$) and parameter $\overline{\Sigma}_{MS}$ (instead of $S_0$) in the fitting procedure in the flicker spectroscopy method.  
%---------------------------------------------------------

Let us summarize the main result of this section in the generalized case when the self consistent is treated numerically. Its solution depends only on  $\overline{C}$ and $\overline{\sigma_0}$. Thus, our goal is to obtain the meanings of $K_c$  and $K_s$ having from  the fitting procedure, on the base of the functional dependence Eq.~\eqref{e670}, the values of $\overline{C}$,  $\gamma$ and $\overline{\sigma_0}$. Using Eqs.\eqref{e1s6} and \eqref{C} we have two possibilities to solve this
problem:
 a) approximative and b) exact.

a)~From Eqs.\eqref{C}) and \eqref{g1} one gets the results:
\begin{equation}
K_s\approx \frac{1}{8\pi\beta R_0^2}\frac{\overline{C}}{\gamma^2}
\label{app}
\end{equation}
and
%\fbox{\begin{minipage}{20em}
\begin{equation}
K_c=\frac{1}{8\pi\beta}\frac{1}{\gamma}.
\end{equation}
%\end{minipage}
It is evident that due to the used approximation  $4\pi R^2_0/S_0 \approx 1$ in obtaining the last term in Eq.~\eqref{C} there is no need to
use the fitting results for $\overline{\sigma_0}$ in order to obtain $K_s$ and $K_c$. In this case Eq.~\eqref{e1s6} has to serve as an estimation of the used approximation if it is rewritten in the form:
\begin{equation}
\frac{4\pi R_0^2}{S_0}-1=\frac{1}{8\pi\beta R_0^{2}}\frac{\overline{\sigma}_0}{\gamma K_s}.
\label{1}
\end{equation}
In other words the above approximation is correct if the rhs of Eq.~\eqref{1}
is $<<1$.

b)~Indeed, the above problem may be solved without any approximation. From the three equations~\eqref{e1s6},\eqref{C} and \eqref{g1} one can obtain the solution in terms of the three functions $K_c=K_c(\gamma),~ K_s=K_s(\gamma,\overline{\sigma}_0,\overline{C})$ and $S_0=S_0(\gamma,\overline{\sigma}_0,\overline{C})$.
Let us exclude the quantity 
\begin{equation}
\frac{4\pi R_0^2}{S_0}\nonumber\\
\end{equation}
from the equations \eqref{1} and \eqref{C}. One gets
\begin{equation}
K_s=\frac{\overline{C}}{8\pi\beta R_0^{2}\gamma^2}\left(1+\frac{1}{8\pi\beta R_0^{2}}\frac{\overline{\sigma}_0}{\gamma K_s}\right)^{-1}.
\label{2e}
\end{equation}
If we skip the second term in the brackets we obtain the approximative result Eq.~\eqref{app}.  From Eq.~\eqref{2e} one obtains: 

\begin{equation}
K_s=\frac{1}{8\pi\beta R_0^2}\frac{\overline{C}}{\gamma^2}\left(1-\frac{\overline{\sigma}_0\gamma}{\overline{C}}\right)
\label{ex2}
\end{equation}

Recall that $\overline {\sigma}_0<0$ and the approximative result for Eq.~\eqref{app} underestimated the exact one, Eq.\eqref{ex2}. The interesting formula relating $K_c$ to the area compressibility modulus $K_s$ to $K_c$  is: 
\begin{equation}
K_s=\frac{K_c}{R_0^2}\left(\frac{\overline{C}}{\gamma}-\overline{\sigma}_0\right).
\end{equation}
This relation should allow one to compare  the relation between $K_s$ and $K_c$ obtained in our approach and the corresponding relation obtained in the framework of other methods (see e.g.  \cite{N13} and refs. therein). 

%------------------------------------------------------------

%\end{document}

\section{The closeness of the model Hamiltonian to the approximating Hamiltonian}

The validity of our method can be controlled  by calculating the  mean square fluctuations of ${\cal A}(\mathrm{v})$ in the upper bound of the Bogoliubov inequalities Eq.~\eqref{e148}, taking into account Eq.\eqref{e149}, defined as
\begin{equation} 
\textit{C}(\widetilde X)\equiv\langle [{\cal A}(\mathrm{v})-\widetilde{X}]^2
\rangle_{H_{app}}(\mathrm{v},\widetilde{X}).
\end{equation} 
From Eqs.~\eqref{eq51} and ~\eqref{e52} it follows that:
\begin{equation}
\widetilde{X}\equiv \langle {\cal A}(\mathrm{v}) \rangle_{H_{app}(\mathrm{v},\widetilde{X})}. \label{e721}
\end{equation}
Obviously,  the correlator $\textit{C}(\widetilde X)$   may be presented  as:
\begin{eqnarray}
&&\langle [{\cal A}(\mathrm{v})-\langle {\cal A}(\mathrm{v}) \rangle_{H_{app}(\mathrm{v},\widetilde{X})}]^2
\rangle_{H_{app}(\mathrm{v},\widetilde{X})} =\nonumber\\
&& \langle [{\cal A}(\mathrm{v})]^2
\rangle_{H_{app}(\mathrm{v},\widetilde{X})} 
\mbox{}-  \langle
{\cal A}(\mathrm{v}) \rangle_{H_{app}(\mathrm{v},\widetilde{X})}^2.
\end{eqnarray}
From Eqs.~\eqref{e39}, \eqref{e1s6}, and \eqref{e670} we obtain:
\begin{eqnarray}
\langle {\cal A}(\mathrm{v})\rangle_{H_{app}(\mathrm{v},\widetilde{X})}
=\frac{\sigma_1}{4\beta}\sum_{n=2}^{n_{max}}
\frac{2n+1}{[n(n+1)+\overline{\Sigma}_{app}]}.  \label{e72}
\end{eqnarray}
From the other side  Eqs.~\eqref{e18} and \eqref{e39} imply:
\begin{eqnarray}
&&\langle [{\cal A}(\mathrm{v})]^2\rangle_{H_{app}(\mathrm{v},\widetilde{X})} =\frac{K_s}{2S_0}
\frac{R_0^4}{4} \nonumber \\
&&\times \sum_{n=2}^{n_{max}} \sum_{m=-n}^{n}
\sum_{n'=2}^{n_{max}} \sum_{m'=-n'}^{n'} (n-1)(n+2) \nonumber \\ &&\times
(n'-1)(n'+2) \langle(v_n^m)^2(v_{n'}^{m'})^2
\rangle_{H_{app}(\mathrm{v},\widetilde{X})}.
\end{eqnarray}
Taking into account that the amplitudes $v_n^m$ are not correlated (the
approximating Hamiltonian presents a system of not interacting oscillators)
and have a Gaussian distribution, we obtain that:
\begin{equation}
\langle (v_n^m)^4\rangle_{H_{app}(\mathrm{v},\widetilde{X})} =3[\langle
(v_n^m)^2\rangle_{H_{app}(\mathrm{v},\widetilde{X})}]^2.
\end{equation}
After some tedious but simple calculations we get:
\begin{eqnarray}
&&C(\widetilde X)=\langle [{\cal A}(\mathrm{v})-\widetilde{X}]^2
\rangle_{H_{app}(\mathrm{v},\widetilde{X})} 
= \nonumber\\
&&\frac{K_s}{S_0} \frac{R_0^4}{4}
\Bigg[\frac{1}{\beta K_c} \Bigg]^2 \sum_{n=2}^{n_{max}}
\frac{2n+1}{[n(n+1)+\overline{\Sigma}_{app}]^2} \label{e661}.
\end{eqnarray}
In the above expression $\overline{\Sigma}_{app}$ is the solution of the
self-consistent equation Eq.~\eqref{e68c} at fixed $kT$, $K_c$,$K_s$, $R_0$, and $S_0$.
When $\overline{\Sigma}_{app} \to -6$  the correlator $C(\widetilde X)$ 
diverges and the estimation becomes useless.
However, one should always keep in mind that in this limit  the contribution of the neglected higher order terms in the expansions Eqs.~\eqref{gS} - \eqref{he11} increases and one would require an expansion in Eqs.~\eqref{gS} - \eqref{he11} beyond the quadratic terms \cite{S95}.

If the correlator is a small quantity in some sense (or equals zero),  then due to the inequalities, Eq.~\eqref{e148}, the
thermodynamics of the model system,
Eq.~\eqref{e1z}, is well approximated (some times colled thermodynamically equivalent) by the approximating
Hamiltonian $H_{app}(\mathrm{v},\widetilde{X})$.

It is instructive to consider the behavior of the correlator Eq.~\eqref{e661} as a function of $K_s$ at the extreme values $0$ and $\infty$.
When  $K_s \rightarrow 0$, from Eq.~\eqref{bt3} it follows that at fixed $kT$,
$K_c$, $R_0$, and $S_0$, the correlator in Eq.~\eqref{e661} also tends to
zero. When $K_s \rightarrow \infty$,
$\overline{\Sigma}_{app}$ tends to $\overline{\Sigma}_{MS}$ (see Eq.~\eqref{bt4}), and the correlator tends to $\infty$.

The sum in the rhs of the Eq.(\ref{e661}) has an asymptotic behavior  in $N$ given by Eq.(\ref{corr2}) (see the Appendix A) in which $\overline{\Sigma}_{app}$ must be replaced with
its value from Eq.(\ref{007}) or Eq.(\ref{Lso18}). As a result, it is easily seen that if $N\to\infty$ then $\textit{C}(\widetilde X)\to 0$ and our calculations are asymptotically exact in the thermodynamic limit $\frac{N}{V}=\mathrm{const}.$

Since we discuss the role of the membrane stretching elasticity, we need to know the validity of our approach as  function of $K_s$. Here it is the place to note that the attempt to calculate even numerically the free energy in conjunction with the self-consistent equation may turn out a rather cumbersome  task. A more efficient way of solving the problem, which avoids the numerical solution of the self-consistent  equation, is to take into account the inverted form of the relation between $\overline{\Sigma}_{app}$ and $K_s$ as given by Eq.(\ref{Ks}).
In other words due to the specific form of this relation it is more convenient instead of $K_s$ to use as an open parameter  $\overline{\Sigma}_{app}$.  To this end	we 		
substitute the variable $\widetilde{X}$ by $\overline{\Sigma}_{app}$ in the Bogoliubov inequalities~\eqref{e46} using the relation (\ref{d83}). Thus, the Bogoliubov inequalities may be rewritten in the form:

\begin{eqnarray}
&& 0\le \frac{ f[H] - f[H_{app}(\overline{\Sigma}_{app})]}{|f[H_{app}(\overline{\Sigma}_{app})]|}\le R(\overline{\Sigma}_{app}),
\label{e46a}
\end{eqnarray}
where
\begin{equation}
R(\overline{\Sigma}_{app})\equiv\frac{C(\overline{\Sigma}_{app})}{|f[H_{app}(\overline{\Sigma}_{app})]|}
\label{re}
\end{equation}
is the relative error.
The behavior of the function $R(\overline{\Sigma}_{app})$ for some fixed $\overline{\Sigma}_{MS}$ is studied in \cite{BT19}.
Our numerical analysis 
shows  that $R(\overline{\Sigma}_{app})<<1$, and therefore the used approximation  provides a very good relative accuracy for any solution $\overline{\Sigma}_{app}$  of the self-consistent equation which belongs to the {\it open } interval $ (-6,\infty)$.
%\end{document}

\section{Outlook of the method considered in Sec.X}

So far, we restrict ourself to the case when the vesicle membrane is a {\it compressible 2D monolayer} immersed in fluids having the same viscosity on the either side of the membrane. 
However actually the effects of interlayer coupling in fluctuating {\it bilayer}  membrane is of  strong interest. A review of some experimental and theoretical results, that have played seminal roles in the field, the reader can found in \cite{ERS13,MM15}. Especially, theoretical descriptions in terms of the discrete spherical harmonics  have been investigated intensively in refs.~\cite{YE95,B99,B10,Mlk02,Kok16}.
Further, investigations of bilayer structures involving interconnected effects
of non-linear  area-elasticity and relative displacement  of the membrane monolayers would be of undoubted interest, therefore, some ideas and problems will be discussed below. 

An important  consequence of the membrane bilayer structure is
that bending deformation is always accompanied with stretching of one monolayer (the outer) and compression of the other (the inner).

It is thus desirable  to utilize the AHM
considered in Sec.10 to the study of the thermal fluctuations of a such more complex bilayer systems. As experience from other fields of condensed matter physics shows it is generally clear that any non-local term added in the Hamiltonian can be treated in this way \cite{B00,B81,B84,Bj13}.
The result would be the appearance of an additional  variation parameter satisfying the corresponding self-consistent equation.

A straightforward generalization of the model Hamiltonian Eq.~\eqref{Hs1a} is to add the term due to the
the relative elastically  expansion of the two individual membrane monolayers. In this model, each monolayer has a preferred (or relaxed) area $S_0^{in}$ and $S_0^{out}$, based on the number of lipid molecules it contains and can have the corresponding actual area $S_0^{in}$ and $S_0^{out}$, respectively. As a result an area difference elasticity (ADE) Hamiltonian \cite{SBZ85,MSWD94} between the two monolayers  may be conveniently expressed in the form:
\begin{equation}
H_{r}=\frac{1}{2}\frac{K_{\Delta}}{S_0}\Bigg(\Delta S({\mathrm{v}})-\Delta S_0\Bigg)^2
\label{ADE}
\end{equation}
where $K_{\Delta}$ is the appropriate elastic constant (non-local bending modulus) and for the term in the denominator the assumption  $S_0^{out}\simeq  S_0^{in}\simeq S_0$ is
used. The term in the brackets,  in
spherical harmonics presentation up to  the second-order approximation in the amplitudes $(v_n^m)$, takes the form \cite{S10}
\begin{eqnarray}
&&\Delta S({\mathrm{v}})\equiv
S^{out}({\mathrm{v}})- S^{in}({\mathrm{v}})=8\pi R_0h\times \nonumber\\
&&\left\{
1+ \frac{1}{8\pi}\Bigg[ \sum_{n=2}^{n_{max}}\sum_{m=-n}^n
(n-1)(n+2) (v_n^m)^2 \Bigg]\right\},
\end{eqnarray}
where  $2h$ is the
separation  between the  two monolayers ) and $\Delta S_0\equiv S_0^{out}-S_0^{in}$. If one introduces the tension due to relative area difference
\begin{equation}
\sigma_{\Delta}=\frac{K_{\Delta}}{S_0}\Bigg(\Delta S({\mathrm{v}})-\Delta S_0\Bigg),
\end{equation}
Eq.~\eqref{ADE} may be rewritten in the well known form
\begin{equation}
H_{r}(\mathrm{v})=\frac{S_0}{2K_{\Delta}}(\sigma_{\Delta}(\mathrm{v}
))^2.
\end{equation}
and has to be added in Eq.~\eqref{1IB1}:
\begin{equation}
H(\mathrm{v})=H_c(\mathrm{v})+H_s(\mathrm{v})+H_{r}(\mathrm{v}). \label{1IB2}
\end{equation}

Due to their uniform structure, the last two terms can be treated  using the approximating Hamiltonian method developed in the Sections 9 and 10.

If one  takes into account the bilayer structure of the membrane  the role of the local lipid densities on each monolayer have to be scrutinized  in the theory. It is well recognized that when the bilayer  fluctuates the significant impact have physical  processes that are result of the change in the local  monolayer  densities. The latter one  can be brought about by  the lateral flows of the lipid molecules. 
Thus, new dynamical degrees of freedom related with the lipid density difference between the two monolayers taking into account the quasi-spherical geometry of the membranes have been incorporated in the theory more or less on a phenomenological level \cite{B10,SL93,YE95,B99,Mlk02}, or on the base of some fundamental principles \cite{Mlk02,Kok16}. The problem is how to consider the transverse deformations with respect to an equilibrium reference configuration followed by a lateral redistribution of the molecules within the bilayer, namely flip-flop motions, and the effects of the  intermonolayer friction.
Here, it is not our aim to extend our theory on the case of bilayers. Rather, we give hint that it is possible  pointing out the problematic items that should be solved in a such theory.

A quantitative theory describing
the out-of-plane fluctuations of a {\it flat}  membrane, taking into account  the intermonolayer  friction and two-dimensional viscosity has been developed in \cite{SL93}. Explicit relations for the fluctuations  of the form of a  quasi-spherical vesicle, influenced by the mutual  displacements of the monolayers, comprising its bilayer, for arbitrary values of the fluctuation wave vector, have been obtained in \cite{YE95}. Later it was proved \cite{B99} that in the case of a bilayer membrane, the bending elasticity, participated in the theoretical results is that of a {\it free flip-flop}.
The above result was obtained by taking into account the lateral displacement of the monolayers. Both theories \cite{YE95,B99} reproduce in
form the result of Milner and Safran, see Eq.~\eqref{MSf} in the present study, for the mean square values of the amplitude $u(\theta,\varphi,t)$, indeed, with more rich physical meaning of the corresponding effective bending elasticity modulus and effective surface tension. In order to be more concrete the comparison of Eq.~\eqref{MSf} with the result obtained in \cite{B99} shows that $K_c$ and $\sigma$ must to be replaced by
the free flip-flop bending elasticity $K^{fr}_c$ and $\sigma +\epsilon$, respectively. The crucial point is the obtained  dependence  of   $\epsilon$  on the free flip-flop bending elasticity $K^{fr}_c$, blocked flip-flop bending elasticity $K^{bl}_c$, and a function which is defined through
an equation contained the difference between  the molecular surface densities of the outer and inner monolayers and the flip-flop coefficient
$\xi$, see Eq.~(29) in \cite{B99}. Since the calculations in the above theories are essentially based on a Gaussian theory of fluctuation this is a hint that one may consider an extension to include the variation of the local density variations in the part given by Eq.\eqref{e140} of our model Hamiltonian, Eq.~\eqref{e141}, simply using the above formulated displacement as a mnemonic rule. A couple of remarks concerning the contributions of the above replacements on the excess area are due here. If we would like
to speculate, using Eq.~\eqref{so232}, whether values of $\Delta(\overline{\Sigma}_{app})$  might legitimately be larger or smaller from $\Delta(\overline{\Sigma}_{MS})$ the first we need is to have an estimation of the difference between the values of $K^{fr}_c$ and $K^{bl}_c$. The second remark concerns the solution $\overline{\Sigma}_{app}$. It must be obtained in a self-consistent 
way. The solutions of the former and the last problems are a difficult task.

A step in this direction,
however beyond a self-consistent theory, has been done in refs. \cite{B99,B10}, where the effects related with the {\it stretching elasticity} of the bilayer in conjunction  with lateral monolayer displacement in fluctuating nearly spherical vesicle have been considered. In these works, however a Milner and Safran type of mean-field approximation that the fluctuations of the effective tension {\it are not correlated} with the fluctuations of the  amplitude $u(\theta,\varphi,t)$ has been used. As a result the correlation between $u(\theta,\varphi,t)$ and the surface tension has been lost resulting in an inability to determine the stretching elasticity modulus $K_s$ from the flicker - noise analysis experiments.

Actually, a consistent approach based on some fundamental principles, have to be done in the framework of the theory proposed in \cite{Mlk02} involving however  the {\it non-linear} elasticity energy of the bilayer, i.e. term of the type Eq.~\eqref{Hs1a}.

Though our method based on the Bogoliubov inequalities  is more generally applicable, in this case various less than trivial problems need to be solved. First, an inevitable issue is the justification of the appropriate choice of the effective Hamiltonian governing the elastic properties of the bilayer. Here, the obstacle is the appropriate choice
of the physical parameters and the corresponding reference states entering in the definition of the Hamiltonian in order to make
relations with the experiment (see e.g.~the "second remark" in ref.~\cite{Mlk02} about the involving a {\it nonlinear} area elasticity).

To include the local density variations in the two monolayer halves and the corresponding  functional measure over an appropriate set of independent degree of freedom two more fields
are needed in addition to $u(\theta,\varphi,t)$ (in our notations  to $v(\theta,\varphi,t)$): $\phi^{+}(\theta,\varphi,t)$) and
$\phi^{-}(\theta,\varphi,t)$, representing the local surface (number) densities 
of the outer and inner monolayer, respectively,
and defined with respect to the surface described by
${\cal R}(\theta,\varphi,t)$. 
As it was pointed out in \cite{Mlk02} the choice of the
set of independent degrees of freedom in the corresponding expression of the Hamiltonian is a sophisticated problem, if the  lateral  flows  of lipid  molecules  must to be taken into account. This might be  a part of the general and complicated problem of the correct construction of statistical ensembles of surfaces \cite{CLNP94}.
These are the necessary points to be clarified in order  to give correct self-consistent formulas for the free energies and correlation functions in the Bogoliubov variational inequalities, Eq.~\eqref{e46}. Moreover, the very solution of the variational problem will be more complicated. 
It is obvious  that the case of  fluctuating  quasi-sperical bilayer
involving a {\it nonlinear area elasticity} is still waiting for an exact theoretical development.

%\end{document}

.

\section{Summary}

Depending on the geometry, there are two different ways to describe the behaviour  of a thermally fluctuating  surface of a vesicle : 

- out-of-plane fluctuations of a {\it flat} membrane, with periodic boundary conditions using the Monge representation, and 

- shape fluctuations of a {\it closed} nearly spherical membrane, using a series expansion with respect to the spherical  harmonics.

Only the latter case, with its specific features, has been analyzed in this review.

In  most theoretical papers a key problem is how to introduce the volume and surface conservations of the vesicle  from some basic principles.
In the previous sections the area constraint has been considered  within three different scenarios:

i.) in an exact manner  through a delta-function in the partition function, see also \cite{S97,Fap01}, 

ii.) involving Lagrange  multipliers in the Hamiltonian to accomplish constraints for the mean area, see also \cite{MS87,F89,S97,Hnf16}.

iii.) involving an elastic contribution term in the Hamiltonian as  considered in our paper \cite{BT19}, see also \cite{B02,B10} .

Though the importance of the results obtained in the first two scenarios are well evaluated,  we like better the last one,  as  physically most natural and consistent with the statistical mechanics requirements.
Let us briefly summarize the motivations for this statement.

In  Sections 1 and 2 we consider the membrane as incompressible and impermeable. Then the volume and  the area of the vesicle can be considered as constrained in the framework of scenarios i) and ii).
In these cases the computational problem  is the implementation of  the fixed constant area if the volume has been already  fixed. 

Scenario i.) is  considered in Section 3. It is the exact realization of the area constraint  by adding a delta-function in the partition function. 
In this case calculations are based on
the  method of the steepest descent  which requires  skills  in  complex analysis in order to prove the existence of the solutions. Last but not least this approach  becomes  exactly valid in the thermodynamic limit. The existence of a thermodynamic limit is an important ingredient of the theory since the thermodynamic ensembles become equivalent only in this limit. Here the problem is that if the membrane is in equilibrium the very existence of the thermodynamic limit becomes questionable and needs to be scrutinized (see Section 5).

Scenario ii.) is  considered in Section 4. It  is the simplest realization of the area constraint and  is achieved in the so called conventional approach with an effective tension. 
The membrane area constraint is guaranteed by a Lagrange
multiplier $\sigma$ conjugate to the real area $A(\mathrm{u})$. 
In this case the real area of the membrane is not fixed,  but its average value $\langle A(\mathrm{u})\rangle_{H(\mathrm{u};\sigma)}$ is controlled by the parameter $\sigma$. 
Its value cannot be measured directly,  it is determined through the temperature and the excess area \cite{S95,S97}. 
Contrary to the apparent simplicity of this approach,  the relation of $\sigma$ to the other   generic  definitions 
of membrane surface tension is  a matter of a longstanding debate (see e.g.\cite{Bif10,F11,Sch11,Hnf16} and refs. therein). 

%Тhe advantage of this method is that it
%allows  easier  analytical
%calculations (the corresponding integrals in the partition function %and thermodynamic mean values are Gaussian) in comparison to scenario %i)----------------------------------------.

Scenario iii) is considered in  Sec. 6 and 7, in which the membrane of the vesicle is treated as a stretchable/compressible thin surface whose elastic response depends on its intermolecular forces.	
An instructive question is whether  it is possible to reveal  the microscopic origin of $\sigma $ in the formulas of  Milner and Safran for the mean square amplitudes and excess area. This should allow to include  the experimental  determination of the stretching elasticity modulus $K_s$ in the flicker spectroscopy method. Having this in mind,  the area dilation energy in the Hamiltonian of the fluctuating system should be taken into account as well. 
However, then the corresponding Hamiltonian $H({\mathrm{v}})$ becomes nonlinear with respect to the squares of spherical harmonics amplitudes, appearing in
the expansion of the vesicle shape fluctuations. As a consequence the standard used tool - the equipartition theorem becomes inapplicable.
In order to solve this
problem one can choose between two different paths: 

1.) 
In the commonly used approach the linearization of the computational problem is  based on the Habbard-Stratonovich transformation  with the subsequent use of the saddle-point approximation \cite{Fap01,Hi04,Lli11}. It turns out that
the problem
is  exactly solvable (only) in the thermodynamic limit
\cite{Fap01,Hi04,Lli11,F11}. Let us recall that firstly this aspect of the theory of flat membranes with periodic boundary conditions  has been discussed in the context of  the spherical model of phase transitions in 1976 \cite{Bgp76}.

2.) In the approach developed in \cite{BT19} the linearization of the Hamiltonian in  Eq.~\eqref{z148} is 
based on the Bogoliubov  variational inequalities.	
In our opinion this approach allows easier to  estimate  the used approximation and have not need to use the complex plane analysis. Moreover, the approximation is not necessarily related to
the notion of the thermodynamic limit. The problem is reduced to solving the self-consistent  equation ~\eqref{e68} for the auxiliary variable $X$ in a finite-size system.
At $X=\widetilde{X}$ this equation has a simple physical
interpretation, if it is presented identically in the form:
\begin{eqnarray}
\langle S({\mathrm{v}})\rangle_{H_{app}({\cal V},\widetilde{X})} 
= A(\widetilde{X}),\label{fa}
\end{eqnarray}
where
\begin{equation}
A(\widetilde{X})= 4\pi R^2 + \sqrt{\frac{2S_0}{K_s}}\widetilde{X}.
\end{equation}
Comparing with the argument of the delta-function, Eq.~\eqref{dl},  where the microscopic area of the vesicle  is fixed in an exact manner in the partition function, we see
that the Eq.~\eqref{fa} (valid for membrane parameters: $S_0,K_s, K_c, R$, and temperature $T$) imposes a "soft"  constraint on the  amplitudes of the shape fluctuations of the vesicle.
It assures that the mean area  of the membrane (lhs of Eq.~\eqref{fa}) is equal to the area $A(\widetilde{X})$ (rhs of Eq.~\eqref{fa}).
As a first step, since $X$  has been introduced to linearize the Hamiltonian, Eq.~\eqref{e141}, the quantity  $\overline{\Sigma}_{app}(\widetilde{X})$ no need to be considered as a direct experimentally measurable quantity. Further on, if $\overline{\Sigma}_{app}(\widetilde{X})$ is considered as a fitting  parameter, then Eq.~\eqref{e670} for  $\langle (v_n^m)^2\rangle_{H_{app}({\cal V},X)}$ can be used to determine  the bending elasticity modulus $K_c$. Of course, we may stop here in the interpretation of the  obtained result. In other words, a $\grave{a}~la$  Milner and Safran approach  may be utilized 
also  for vesicles with a compressible thin film membrane. If in this case  the flickering analysis works well enough following the  preexisting conventional approach only means that an experimental situation is realize where the importance of $K_s$  is not that important.

The problem with  approach based on the Lagrange multiplier $\sigma$ is that its physical significance  is somewhat ambiguous in the context of thermal fluctuations. There is no clear reason to believe that the membrane tension is independent of the shape fluctuations, more so if one ignores the role of stretching elasticity of the membrane. There are no such problems with $\overline{\Sigma}_{app}$ which is the membrane tension as follows from Eq.~\eqref{ks1}.
Tough to extract  from the functional expression Eq.~\eqref{e670n} a quantitative information about $K_s$ is not so trivial, this is generally possible. To do this three quantities: $K_c, K_s$ and  $\overline {\Sigma}_{MS}$ have to be inferred as fitting parameters from the flickering analysis of the measurable shape fluctuations of the vesicle. In the general case, described by the functional expression~\eqref{e670}, since the self-consistent equation depends only
on $\overline{\sigma}_0$ and $\overline{C}$, it is convenient to use
as fitting parameters $\gamma, \overline{\sigma}_0$ and $\overline{C}$.

This consideration reveals a possibility to extract the value of the
stretching elasticity modulus $K_s $ in conjunction with the estimation of the exactness of the used approximation. 
The degree of the exactness of the results can be obtained by estimating the correlator in the rhs of the Bogoliubov inequalities (see Eq.~\eqref{e661}). This estimation can be applied for finite membranes as well.

%\end{document}

.

\section{Acknowledgments}

This review is based on my lecture  at the Conference in memory of  Vyatcheslav Borisovich Priezzhev held in the Bogoliubov Laboratory of Theoretical Physics (JINR-Dubna, 10 September 2019). I am  grateful to the Organizing Committee and especially to V.P Spiridonov and A.M. Povolotsky, for the invitation and hospitality

I am grateful to I. Bivas for numerous stimulating discussions on the physics of vesicles concerning theory and experiment. Many of the ideas presented in this review are based on our previous common works.
I would like to thank A.G.Petrov for his useful comments on the manuscript.

This work is partly supported by the  JINR (Dubna)-ISSP-BAS (Bulgaria)  collaborative Grant "Investigation of the influence of nanoparticles on the properties of biologically relevant systems" 2019/2021. 
%\end{document}
\appendix 

\section{Limit case analytical solutions of Eqs.~\eqref{b101} and~\eqref{e68c} }

Eqs.~\eqref{b101},~\eqref{e68c} and   (\ref{e72}) could
be studied analytically  by replacing the sum  in its rhs  with an integral. In order to validate the corresponding approximation we shall use the Euler-McLaurin summation formula
\begin{eqnarray}
&&\sum_{n=0}^{n_{max}-1}F\left(n+\frac{1}{2}\right) = \int_{0}^{n_{max}}F\left(t+\frac{1}{2}\right) dt
-\nonumber\\
&&\frac{1}{2} \left[F\left(n_{max}+\frac{1}{2}\right) -F\left(\frac{1}{2}\right)\right]+\nonumber\\  
&& \frac{1}{12}[F^{'}\left(n_{max}+\frac{1}{2}\right) - F^{'}\left(\frac{1}{2}\right) ] + \dots,
\label{EM}
\end{eqnarray}
where
\begin{equation}
F(x)=F_{1}(x)=\frac{2x}{x^2 + \Sigma-1/4}
\label{f1}
\end{equation}
with  $\Sigma =\overline{\Sigma}_{MS}$ in Eq.(\ref{b101}) and $\Sigma =\overline{\Sigma}_{app}$ in Eq.(\ref{e68c}), and
\begin{equation}
F(x)=F_{2}(x)=\frac{2x}{[x^2 + \Sigma-1/4]^{2}}
\label{f2}
\end{equation}
with $\Sigma =\overline{\Sigma}_{app}$ in Eq.~\eqref{e72}.

Let us ignore: i.) the higher order terms in Eq.~(\ref{EM}), and ii.) approximate $F(x)\approx F(0) + xF^{'}(0)$ in the interval $[0,1/2]$. The approximations made are consistent only for large $\overline{\Sigma}_{app}>>1$ since then the relative change of $F(x)$ is small  when $n \to n+1$. With these approximations the Euler-Maclaurin formula Eq.~(\ref{EM}) reduces to 
\begin{widetext}
\begin{eqnarray}
\sum_{n=0}^{n_{max}}F\left(n+\frac{1}{2}\right)\approx\int_{0}^{n_{max}+\frac{1}{2}}F(x)dx+\frac{1}{24}F^{'}(0)+ 
\frac{1}{2}
\left[F\left(n_{max}+
\frac{1}{2}\right) + \frac{1}{6}F^{'}\left(n_{max}+\frac{1}{2}\right)\right] \label{EMR}
\end{eqnarray} 
\end{widetext}
(c.f. with Eq.~(59.10), p.173 \cite{LL80}). 
Using Eq.~\eqref{EMR} the summation in Eqs.\eqref{b101} and \eqref{e68c}  can be performed easily. The result is 
\begin{eqnarray}
&&\sum_{n=2}^{n_{max}}F_{1}\left(n+\frac{1}{2}\right)\approx\nonumber\\&&	\ln\frac{N+N^{1/2}+\Sigma} {\Sigma-1/4}+\frac{1}{12}\frac{1}{\Sigma-1/4}+\nonumber\\
&&\frac{1}{2}\frac{2N^{1/2}+1}{N+N^{1/2}+\Sigma}-\frac{1}{6}
\frac{N+N^{1/2}-\Sigma+1/2}{(N^2+N^{1/2}+\Sigma)^2}, \nonumber\\
\label{s1}
\end{eqnarray}
and from where it follows
\begin{eqnarray}
&&\sum_{n=2}^{n_{max}}F_{1}\left(n+\frac{1}{2}\right)\approx \nonumber\\
&& \ln{\frac{N}{\Sigma}} +\frac{\Sigma}{N}	
+ O\left(\frac{1}{N^{1/2}}\right)+
O\left(\frac{1}{\Sigma}\right) + O\left(\left[\frac{\Sigma}{N}\right]^2\right), \nonumber\\
\label{s11}
\end{eqnarray}
in the case of	Eqs.~\eqref{b101} and \eqref{e68c}, and
\begin{eqnarray}
&&\sum_{n=2}^{n_{max}}F_{2}\left(n+\frac{1}{2}\right)\approx \nonumber\\	&&\frac{1}{\overline{\Sigma}_{app}}-\frac{1}{\overline{\Sigma}_{app}+N} +O\left(\frac{1}{\overline{\Sigma}^2_{app}}\right),
\label{corr2}
\end{eqnarray}
in the case of Eq.\eqref{e661}. In the above expressions it is used that $n_{max}\approx \sqrt{N}$. 

A.1 Solution of Eq.~\eqref{b101}.

Let us introduce the notation 
\begin{equation}
x_0=-\frac{\overline{\Sigma}_{MS}}{N}.
\label{L0j}
\end{equation}
With the help of Eq.~\eqref{s11} and the definition of $\overline{\sigma}_{0}$ (see Eq.(\ref{e16})),  Eq.(\ref{b101})  may be presented (up to the used approximations) in the form:
\begin{equation}
x_0e^{{x_0}} =-e^{-\frac{\Delta}{\gamma}},
\label{Leqp1}
\end{equation}

Eq.~\eqref{Leqp1}  can be solved in terms of the Lambert function  ${\bf W}(x)$. A review of its mathematical properties and physical applications can be found in \cite{Cor96,C13,KP10,PI13,PI15} and refs. therein.
Recall that by definition 
\begin{equation}
{\bf W}(xe^x)=x. 
\label{dLf}
\end{equation}
The Lambert function  can take two possible real
values for $-\frac{1}{e} \leq x \leq 0 $. Values satisfying ${\bf W}(x)\geq -1$ belong to the principal branch denoted as ${\bf W}_0(x)$, while values satisfying ${\bf W}(x) \leq -1$ belong  to the ${\bf W}_1(x)$ branch.
The two branches meet at the branch point for $x=-\frac{1}{e}$,
where ${\bf W_0}(-\frac{1}{e})={\bf W}_{-1}(-\frac{1}{e})$. All values of ${\bf W}$ for $x \geq 0 $ belong to the principal branch ${\bf W}_0(x)$.  

The solution of Eq.~\eqref{Leqp1} now reads
\begin{equation}
x_0={\bf W} \left(-e^{-\frac{\Delta}{\gamma}}\right),
\end{equation}
or finally 
\begin{equation}
\overline{\Sigma}_{MS}=-N{\bf W} \left(-e^{-\frac{\Delta}{\gamma}}\right).
\label{sar}
\end{equation}
In the interval $ -e^{-1} \leq -e^{-\frac{\Delta}{\gamma}}<0$ the equation has two solutions given by ${\bf W}_0$ and ${\bf W}_{-1}$, respectively. 

For large $x$, the function ${\bf W}(x)$ is approximated by 
\begin{equation}
{\bf W}(x)=\ln x-\ln\ln x +o(1).
\label{lf1}
\end{equation} 

For small $x$, the Taylor series around $x=0$ is given by 
\begin{equation}
{\bf W}(x)=x-x^2+ ...
\label{lf2}
\end{equation}

The first few terms of the series expansion of  ${\bf W}(x)$ near the branching point  are
\begin{equation}
{\bf W}(x)=-1+p -\frac{1}{3}p^2+ ..., 
\end{equation}
where $p=\pm \sqrt{2(e.x+1)}$ for  ${\bf W}(x)_{0,1}.$

Thus, using Eq.~\eqref{lf2} for $x=e^{-\frac{\Delta}{\gamma}}<<1$, one gets  Eq. (\ref{sol02}):
\begin{equation}
\overline{\Sigma}_{MS}=Ne^{-\frac{\Delta}{\gamma}}.
\label{sol01}
\end{equation}

Using the expansion near the branching point of  the Lambert function, i.e. $x= e^{-\frac{\Delta}{\gamma}} \approx e^{-1}$, 
one obtains:
\begin{equation}
\overline{\Sigma}_{MS}=N\left[ 1-\sqrt{2(1-e^{-\frac{\Delta}{\gamma}+1})}\right].
\end{equation}

A.2 Solution of Eq.~\eqref{e68c}

For $\overline{\Sigma}_{app}>>1$, Eq.~\eqref{e68c} can be treated in the same way. Let us introduce the notation
\begin{equation}
x=\left(\frac{1}{\overline{C}}-\frac{1}{N}\right)\overline{\Sigma}_{app}.
\label{L0}
\end{equation}.
Using Eq.~\eqref{s11} the self-consistent equation~\eqref{e68c} may be presented (up to the used approximations) in the form:
\begin{equation}
xe^{x} =\left(\frac{1}{\overline{C}}-\frac{1}{N}\right)
Ne^{\overline{\sigma}_{0}/\overline{C}}.
\label{Leqp}
\end{equation}
In terms of the Lambert ${\bf W}(x)$  function  the solution reads:
\begin{equation}
\overline{\Sigma}_{app}=\left(\frac{1}{\overline{C}}-\frac{1}{N}\right)^{-1}{\bf W}\left[\left(\frac{1}{\overline{C}}-\frac{1}{N}\right) N\exp\left(\frac{\overline{\sigma}_{0}}{\overline{C}}\right)\right].
\label{Lsol}
\end{equation}
Thus, if $\frac{1}{\overline{C}}-\frac{1}{N}<0$ there will be two solutions or none (or only one solution if the argument of {\bf W} is exactly $-\frac{1}{e}$).
If $\frac{1}{\overline{C}}-\frac{1}{N}>0$ there will be one solution. 

With the help of bout expansions the Lambert ${\bf W}(x)$  function (\ref{lf1}) and (\ref{lf2}) one easily  obtains Eqs.(\ref{007}) and (\ref{Lso18}). 

Not that if
\begin{equation}
-\frac{\Delta}{\gamma}=\frac{\overline{\sigma}_0}{\overline{C}},
\end{equation} 
combining Eqs.\eqref{Lsol} and \eqref{Leqp1},
the more general relation takes place
\begin{equation}
\overline{\Sigma}_{app}=\left(\frac{1}{\overline{C}}-\frac{1}{N}\right)^{-1}{\bf W}\left[\left(\frac{1}{\overline{C}}-\frac{1}{N}\right)
\overline{\Sigma}_{MS}\exp\left(-\frac{\overline{\Sigma}_{MS}}{N}\right)\right].
\label{Lwol}
\end{equation}
From the above result, if $K_s\to \infty$, immediately follows Eq.~\eqref{bt4} where use has been made of  definitions \eqref{C} and \eqref{dLf}.

%\end{document}

%\appendix 
\section{the Griffits-Fisher lemma}

There is a mathematical statement known as  Griffits-Fisher lemma \cite{G64,F65}, which asserts  that if a sequence of {\it convex} function converges pointwise to a limit function, then  the sequence of its derivatives converges to the derivative of the limit function at the  points of its continuous differentiabilty. More precisely, if all functions $\{f_n(x)\}$ and the limit function $f_{\infty}(x)$ are differentiable at a point $x_0\in I \subset R$, then
\begin{equation}
\lim _{n \to \infty} f'_n(x_0)= f'_{\infty}(x_0).
\end{equation} 
More general result due to Fisher consider the case of non-differentiable functions with left and right derivatives at any point  $x\in I$. The latter is relevant if the systems undergo  thermodynamic phase transitions with spontaneously symmetry breaking. These statements are useful in proving the asymptotic closeness of certain average values in the model and approximating system, see e.g. \cite{B00}.
In our case, we consider  both Hamiltonian $H$, Eq.~\eqref{e1z}, and
$H_{app}$, Eq.~\eqref{e2}, and introduce the following  auxiliary
Hamiltonians 
\begin{equation}
{\cal H}(h)=H + h\sigma(\mathrm{v})\label{aH}
\end{equation}
and
\begin{equation}
{\cal H}_{app}(h)=H_{app} + h\sigma(\mathrm{v}),
\label{aH1}
\end{equation}
where $h$ is an auxiliary real parameter which at the end of the calculations will sent to zero.
Further, we obtain that
\begin{equation}
\langle \sigma(\mathrm{v})\rangle_{H}=\frac{\partial}{\partial h}F[{\cal H}(h)]\Bigg |_{h=0}
\end{equation}
and

\begin{equation}
\langle \sigma(\mathrm{v})\rangle_{H_{app}}
=\frac{\partial}{\partial h}F[{\cal H}_{app}(h)]\Bigg |_{h=0}
\end{equation}

%A natural assumption is the existence of a limit (for example %thermodynamic) of the free energy (density) for the auxiliary
%Hamiltonian Eq.~\eqref{aH}. Denoted as
%\begin{equation}
%	\lim_{\cal N \to \infty}\frac{F(H)}{{\cal N}} <\infty
%	\end{equation}
%the limit function (the free energy of the membrane is extensive). 
In the limit when the analog of the correlator Eq.~\eqref{e661} in the rhs of the Bogoliubov inequalities with Hamiltonians \eqref{aH} and \eqref{aH1} tends to zero as a function of its parameters:
\begin{equation}
F[{\cal H}_{app}(h)] \to F[{\cal H}(h)],
\end{equation}
Since $F[{\cal H}_{app}(h)]$ and $F[{\cal H}(h)]$ are convex differentiable functions of $h$ from the lemma follows that, 

\begin{equation}
\langle \sigma(\mathrm{v})\rangle_{H_{app}} \to \langle \sigma(\mathrm{v})\rangle_{H}.
\end{equation}
For the above proof  to be  correct
the definition of the thermodynamic (or other) limit  should be scrutinized.

%\end{document}


\begin{thebibliography}{99}
%\def\selectlanguageifdefined#1{
%	\expandafter\ifx\csname date#1\endcsname\relax
%	\else\selectlanguage{#1}\fi}
%\providecommand*{\href}[2]{{\small #2}}
%\providecommand*{\url}[1]{{\small #1}}
%\providecommand*{\BibUrl}[1]{\url{#1}}
%\providecommand{\BibAnnote}[1]{}
%\providecommand*{\BibEmph}[1]{\emph{#1}}
%\ProvideTextCommandDefault{\cyrdash}{\hbox to.8em{--\hss--}}
%\providecommand*{\BibDash}{\ifdim\lastskip>0pt\unskip\nobreak\hskip.2em\fi
%	\cyrdash\hskip.2em\ignorespaces}

\bibitem{Sin72}
Singer  J. S and  Nicolson G. L. "The fluid mosaic model of the structure of cell membranes", Science, (1972) v.~175, no.~4023.
 pp.~720--731.
 
 \bibitem{P99}
  {Petrov A.G.} "The Lyotropic State of Matter: Molecular Physics and Living Matter Physics",  Gordon and Breach Science Publishers, (1999).
 
 \bibitem{Am17}
 {C. G.Siontorou,  G. P. Nikoleli, D. P. Nikolelis, and  S. K. Karapetis.} "Artificial Lipid Membranes: Past, Present, and Future", Membranes (Basel)
(2017)  v.~7(3), no.~38. 
 pp.~1-99. %\BibDash
 
 \bibitem{S03}
 {Safran S.A.} "Statistical Thermodynamics
 	of Surfaces, Interfaces, and Membranes", Frontiers in Physics. Taylor and Francis Group, (2003).
 	
 	\bibitem{H73}
 	{Helfrich W.} "Elastic properties of lipid bilayers: theory and possible experiments", Z. Naturforsch. C,
 	(1974) 
 	v.~28, no.~11. 
 	pp.~693-703.
 


\bibitem{C70}
{Canham P.B.} "The minimum energy of bending as a posible explanation of the biconcave shape of human red blood  cell"
,J. Theor. Biol.
(1970)
v.~26, no.1. 
pp.~61--81.


\bibitem{E74}
{Evans E. A.} "Bending resistance and chemically induced moments in membane bilayers", Biophys.J.
(1974) v.~14, no.~2. 
pp.~923--931. %\BibDash


 

\bibitem{H86}
{Helfrich W.} "Size distributions of vesicles: the role of the effective
	rigidity of membranes", {J. Phys. (France)}
(1986) 
v.~47, no.~2. 
pp.~321--329.


\bibitem{MSWD94}
{Miao L., Seifert U., Wortis M., and D\"{o}bereinert  H-G. } "Budding transitions of fluid-bilayer vesicles: The efFect of area-difFerence elasticity
", Phys. Rev. A
(1984)
v.~49, no.~6. 
pp.~5389--5407.



\bibitem{D15}
{Deserno M.} "Fluid lipid membranes: From differential geometry to curvature stresses", Chem. and Phys. of Licuids
(2015) v.~185,  
pp.~11--45.



\bibitem{L04}
{Libler S.} "Equilibrium statistical mechanics of fluctuating films and membranes", Statistical mechanics of membranes and surfaces, Ed. by D.~Nelson,T.~Piran, S.Weinberg. (2004)
pp.~49-102.


\bibitem{MS87}
{ Milner S.T. and Safran S.A. } "Dynamical fluctuations of droplet microemulsions and vesicles",
{Phys. Rev. A}
(1987)
v.~36, no.~1. 
pp.~4371--4379.


\bibitem{S95}
{Seifert U.} "The concept of effective tension for fluctuating vesicles",
{Z. Phys. B}
(1995) v.~97,  pp.~299--309. 

\bibitem{S94}
{Seifert U.} "Fluid membranes:
	Theory of vesicle conformations",
{Habilitation theses Ludwlq-Maximllians-Unlversitat, Munchen, Ch.4.2.2}
(1994)


\bibitem{BL75}
{Brochard F. and Lenon J.F.} "Frequency spectrum of flicker phenomenon in erythrocytes"
(1975) v.~36, no.~1. pp.~1035--1047.

\bibitem{SJW84}
{ Schneider M.B., Jenkins J.R. and Webb W.W. } "Thermal fluctuations of large quasi-spherical bimolecular phospholipid vesicles"
{J.Phys.(France)}
(1984)
v.~45, no.~1. pp.~1457-1472--4379.

\bibitem{B87}
{ Bivas I., Hanusse, Bothorel P., Lalanne J. and Aguerre-Chariol O. } "An application of the optical microscopy to the determination of the curvature elastic modulus of biological and model membranes"
{ J. Phys. II}
(1987) v~48, no.~5. pp.~855--867. 


\bibitem{F89}
{Faucon J. F. ,  Mitov M. D., Méléard  P.,  Bivas I. and  Bothorel P.} "Bending elasticity and thermal fluctuations of lipid membranes. Theoretical  and experimental requirements",
{J. Phys. (Paris)}
(1989) v.~50, no.~6. 
pp.~2389--2414.

\bibitem{Mel97}
{Meleard P., Gerbeaud C., Pott T., Fernandez-Puente L., Bivas I, Mitov M.D., Dufourcq J.} "Bending elasticities of modified  membranes:  Influences of  temperature     and  sterol  content ",
(1997) v.~72, no.~6. pp.~2616--2629.


\bibitem{PDPJB04}
{ Pécréaux J., D\"{o}bereiner  H.-G.,  Prost J.,  Joanny J.-F., and  Bassereau P.}" Refined contour analysis of giant unilamellar vesicles"
(2004) v.~13, pp.~277–290.

\bibitem{GVB13}
{Genova J.,Vitkova V. and Bivas I.} "Registration and analysis of shape fluctuations of nearly spherical lipid vesicles"
(2013) v.~88, pp.~022707--(1-9)).



\bibitem{G13}
{Genova J.} {Marin Mitov Lectures}: "Measuring and bending elasticity of lipid bilayer"
{Adv. in Planar Lipid Bilayers and Liposomes, A tribute to Marin D.Mitov}
(2013) v.~17, Ch.~1. 
pp.~1-27. 



\bibitem{VP13}
{Vitkova V. and Petrov A.G.} "Lipid bilayers and membranes: material properties"
{Adv. in Planar Lipid Bilayers and Liposomes, A tribute to Marin D.Mitov }
(2013) v.~17, Ch.~5. pp.~89-138.

\bibitem{MS16}
{Monzel C. and K Sengupta K.
} "Measuring shape fluctuations in biological membranes"
{J. Phys. D: Appl. Phys.}
(2016) v.~49, 
pp.~2430002 (21). 



\bibitem{R17}
{ Rautu S. A.,  Orsi D.,  Di Michele L.,  Rowlands G.,  Cicuta P.and Turner M.S.}"The role of optical projection in the analysis of membrane fluctuations",
{Soft Matter}
(2017) v.~13, no.~19, pp.~3480--3483.

\bibitem{Oh89}
{Ou-Yang Z.C. and Helfrich W.} "Bending energy of vesicle
	membranes:general expressions for the first, second and third variation of the shape energy and application to spheres and cylinders"
{Phys. Rev. A}
(1989) v.~39, pp.~5280--5288.

\bibitem{H92}
{Heinrich V., M. Brumen M., R. Heinrich R. , Svetina S.
	and B. \v{Z}ek\v{s} B.} "Bending energy of vesicle
	membranes:general expressions for the first, second and third variation of the shape energy and application to spheres and cylinders"
{Journal
	de Physique II, EDP Sciences}
(1992) v.~2, no.~5. pp.~1081--1108.% \BibDash	

\bibitem{S10}
{Sev\v{s}ek F.} "Membrane elasticity from shape
	fluctuations of phospholipid vesicles"
{Adv. in Planar Lipid Bilayers and Liposomes }
(2010) v.~12, Ch.~1. 
pp.~1-19.

\bibitem{KS93}
{Komura S. and Seki  K.} "Dynamical fluctuations of spherically closed fluid
	membranes"
{Physica A}
(1993) v.~192, 
pp.~27--46.

\bibitem{Bif10}
{ Barbetta  C.,  Imparato A. and  Fournier J. B.} "On the surface tension of fluctuating quasi-spherical vesicles"
{Eur. Phys. J. E}
(2010) v.~31, no.~3, pp.~333--342. 




\bibitem{GDM17}
{ Gueguen G., Destanville  N. and  Manghi M.} "Fluctuation tension and shape transition of vesicles: renormalisation calculations and Monte Carlo simulations"
{Soft Matter}
(2017) v.~84, no.~5. pp.~6100 (8).




\bibitem{Gk96}
 {Gomper G. and  Kroll D.M.} "Random surface discretizations and the
	renormalization of the bending rigidity"
{J. Phys. I France}
(1996) v.~6, no.~10. pp.~1305--1320.

\bibitem{CLNP94}
 { Cai W, Lubensky T.C., Nelson P , Powers T} "Measure factors, tension and correlations of fluid membranes"
{J. Phys. I France}
(1994) v.~4, 
pp.~931--949. 




\bibitem{D04}
{David F.} "Geometry and field theory of randome  surfaces and membranes", in: Statistical Mechanics of Membranes and Surfaces~/ Ed. by D.~Nelson,T.~Piran, S.Weinberg. (2004) pp.~149-209.

\bibitem{Bi92}
{ Bivas I., Bivolarski L., Mitov  M. and Derzhanski  A.} "Correlations between the form fluctuations modes of flaccid quasi-spherical  lipid vesicles and their role in the calculation of the curvature elastic modulus of the vesice membrane. Numerical results"
{ J. Phys. II}
 (1992) v.~2, no.~7. pp.~1423--1438. 







\bibitem{Fap01}
{ Fournier J.B.,  Ajdari A. and  Peliti L.
} "Effective-area elasticity and tension of micromanipulated membranes"
{Phys. Rev. Lett.}
(2001) v.~86, no.~21. pp.~4970--4973. 


\bibitem{S97}
{Seifert U.} "Configurations of fluid membranes and vesicles"
{Adv. Phys.}
(1997) v.~46, no.~1. 
pp.~13--137. 






\bibitem{J72}
{ Joyce G.S.} Critical properties of the spherical model~// Phase Transitions and Critical Phenomena~/ Ed. by C.~Domb and N.S. ~Green. (1972) v. 2. p.~375.



\bibitem{B00}
{ Brankov J. D., Danchev  D. M., and  Tonchev N. S.} "Theory
	of Critical Phenomena in Finite-Size Systems: Scaling and
	Quantum Effects" ,  World Scientific, Singapore, 2000.

\bibitem{F11}
{ Farago O.} "Mechanical surface tension governs membrane thermal fluctuations"
{Phys. Rev. E}
 (2011) 
v.~84, no.~5. pp.~051914--numpages 8. 


\bibitem{Hnf16}
{ Shiba H., Noguchi H. and  Fournier J. B.} "Monte Carlo study of the frame, fluctuation and internal tensions of fluctuating membranes with fixed area"
{Soft Matter}
(2016) v.~12, no.~8. pp.~2373--2380.

\bibitem{BT19}
{Bivas I. and Tonchev N.S.} " Membrane stretching elasticity and thermal shape fluctuations of nearly spherical lipid vesicles"
{ Phys.Rev.E}
(2019) v.~100, no.~2. pp.~022416(12). 




\bibitem{Hi04}
{ Henriksen J. R. and  Ipsen J. H.} " Measurement of membrane elasticity by micro-pipette aspiration"
{Eur. Phys. J.E (Paris)}
 (2004) v.~14, no.~2. pp.~149-167.

\bibitem{G02}
{Gibbs J. W.} " Elementary Principles in Statistical Mechanics with Especial Reference to the Rational Foundation of Thermodynamics"
,Yale University Press (1902), reprinted by Dover, New York (1960).

\bibitem{A06}
{ Adams S.} "Lectures on mathematical statistical mechanics", Communications of the Dublin Institute for Advanced Studies
Series A (Theoretical Physics), Dublin Institute for Advanced Studies 
(2006).

\bibitem{T15}
{Touchette H.} "Equivalence and nonequivalence of ensembles: Thermodynamic, macrostate, and measure levels"
{J. Stat. Phys.}
(2015) 
v.~159, pp.~987-1016. 

\bibitem{Sch11}
{Schmid F.} "Are strees-free membranes really "tensionles"?"
{Eur.Phys. Lett.}
(2011) v.~95, no.~2. pp.~28008.

\bibitem{Sch13}
{Schmid F.} "Fluctuations in lipid bilayers:Are they understood?"
{Biophys. Rev. Lett.}
(2013)  v.~08, no.~1. pp.~1-20.

\bibitem{Bgp76}
{Brochard F., De Gennes P. G. and  Pfeuty J.} "Surfice tension and deformations of membrane structures:relation to two dimensional phase transitions",
{J.Phys.(Paris)}
(1976) v.~37, no.~10. pp.~1099-1104. 


\bibitem{D97}
{ Marsh D.} "Renormalization of the tension and area expansion modulus in fluid membranes"
{ Biophys.J.}
(1997) v~73, no.~2. pp.~865-869. 



\bibitem{N13}
{ Nagle J.F.} "Introductory lecture: Basic quantities in model biomembranes; Faraday Discuss.",
\href{http://dx.doi.org/10.1134/S1547477114060120}{The Royal Society of Chemistry}
(2013) v.~161,  pp.~11-29. 



\bibitem{Lli11}
{Lamholt  M. A., Loubet B.,  Ipsen J. H.} "Elastic moderation of intrinsically applied tension in lipid membranes",
{Phys. Rev. E }
(2011) v.~83, no.~1. pp.~011913(1-4).

\bibitem{B02}
{ Bivas I.} "Elasticity and shape fluctuation of a lipid membrane",
{Eur. Phys. J. B }
(2002 v.~29, pp.~317-322. 

\bibitem{B10}
{ Bivas I.} "Shape fluctuation of nearly spherical lipid vesicles and emulsion droplets",
{Phys.Rev E }
 (2010) v.~81, no.~6, pp.~061911-1-9. 





\bibitem{Fp03}
{Farago O. and  Pincus P.} "The effect of thermal fluctuation on Schulman area elasticity"
{Eur. Phys. J. E }
(2003) v.~11, no.~4. pp.~399-408.

\bibitem{Fpab13}
{Fosnaric M., Penic S., Iglic and Bivas I.} "Thermal fluctuations of phospholipid vesicles studied by Monte Carlo simulations",
{Adv. in Planar Lipid Bilayers and Liposomes, A tribute to Marin D.Mitov }
 (2013) v~17, Ch.~12. pp.~331-357.
 
\bibitem{Sr86}
{ Shapiro  J. and Rudnick J.} "The fully finite spherical model",
{Phys. Rev. E }
 (1986 v.~43, no.~1/2. pp.~51-83.

\bibitem{MM94}
{Morse D.C. and Milner S.T.} "Fluctuations and phase behavior of fluid membrane vesicles",
{Europhys. Lett. }
(1994) v.~26, no.~8. pp.~565-570. 




\bibitem{L14}
 {Lipowsky R.} "Coupling of bending and stretching deformation in vesicle membranes" 
{Adv. Colloid Interf. Sci.}
( 2014) v.~208, pp.~14-24. 

\bibitem{Z20}
{Zagrebnov V.A.} "Gibbs Semigroups", in: Operator Theory: Advances and Applications, (2019) v. 273,
 Birkh\"{a}user.





\bibitem{B81}
{Bogolyubov (Jr.), Brankov J.G., Zagrebnov V.A., Kurbatov A.M., Tonchev N.S.}"The Approximating Hamiltonian Method in Statistical Physics" (in Russian),  
(1981)  Publ.House Bulg.Acad. Sci, Sofia.



\bibitem{B84}
{Bogolyubov (Jr.), Brankov J.G., Zagrebnov V.A., Kurbatov A.M., Tonchev N.S.} "Some class of exactly soluble models of problems in quantum statistical mechanics", 
{Russian Math. Surveys}
(1984) v.~39, no.~6. pp.~1-50.



\bibitem{Bj13}
{Bogolyubov N.N.(Jr.)} "A Method for Studying Model Hamiltonians: A Minimax Principle for Problems in Statistical Physics", (2013)
 Pergamon.

\bibitem{T67}
{S. V. Tiablikov S. V.} "Methods in Quantum Theory of Magnetism",  (1967) Plenum Press, New York.



\bibitem{AS16}
 {Ahmadpor F. and Sharma  P.} "Thermal fluctuations of vesicles and nonlinear curvature elasticity-implications for size-dependent renormalized bending rigidity and vesilecle size distribution", 
{Soft Matter}
 (2016) v.~12, pp.~2523-2536. 

\bibitem{BT14}
 {Bivas I., Tonchev N.S.} "On the statistical mechanics of shape fluctuations of nearly spherical lipid vesicle", 
{Journal of Physics: Conference Series} (2014) v.558, p.012020   arXiv:1409.3709~[cond-mat].

\bibitem{ERS13}
{Evance E.,Rawicz W. and Smith B.A.} "Concluding remarks
	Back to the future: mechanics and
	thermodynamics of lipid biomembranes"'
{Faraday Discuss.}
(2013) v.~161, pp.~591-611. 

\bibitem{MM15}
{Mell M., Moleiro L. H., Hertle Y. et al.}
"Fluctuation dynamics of bilayer vesicles with intermonolayer sliding: experiment and theory",
{Chem. Phys. Lipids}  (2015) v. 185, pp.~61-77.

\bibitem{YE95}
{Yeung A. and Evance E.} {"Unexpected Dynamics in Shape Fluctuations of Bilayer Vesicles",
{J. Phys.II, France}
 (1995) v.~5, no.~10,  pp.~1501-1523.

\bibitem{B99}
{Bivas I., Meleard P., Mircheva I. and Bothorel} "Thermal shape fluctuations of a quasi-spherical vesicle when the mutual shape fluctuations are taken into account",
{Colloids Surf. A}
 (1999) v.~157,  pp.~21-33.
 
\bibitem{Mlk02}
{Miao L., Lomholt M.A. and Kleis J.} "Dynamics of shape fluctuations of quasi-spherical vesicle revisited",
{Eur. Phys. J. E}
(2002) v.~9, pp.~143-162. 



\bibitem{Kok16}
{ Sachin Krishnan T. V., Okamoto  R., and  Komura S.} "Relaxation dynamics of a compressible bilayer vesicle containing highly viscous fluid",
{Phys. Rev. E}
(2016) v.~94, pp.~062414 (14). 

\bibitem{SBZ85}
{Svetina S., Brumen M. and \u{Z}ek\u{s} B.} "Lipid bilayer elasticity and the bilayer couple interpretation of red cell shape transformations and lysis",
{Stud. Biophys.}
 (1985) v.~110, pp.~177-184. 

\bibitem{SL93}
{Seifert U. and Langer S.A.} "Viscous modes of fluid bilayer membranes",
{Europhys.Lett.}
 (1993) v.~23, pp.~71-76.



\bibitem{LL80}
{ Landau L. D. and  Lifshitz E. M.} "Statistical Physics", 3rd ed. (1980), Pergamon Press, Oxford.

\bibitem{Cor96}
{ Corless R. M., Gonnet G., Jeffrey D., and  Knuthi D. E.} "On the Lambert W Function"
{Adv.Comput. Math.}
(1996) v.~5, pp.~329-360.

\bibitem{C13}
{ Chatzigeorgiou I} "Bounds on the Lambert function and their application
	to the outage analysis of user cooperation",
{IEEE Commun. Lett.}
(2013) v.~17, pp.~1505-1508. 

\bibitem{KP10}
{Kazakova S.G.and  Pisanova E.S.} ""Some Applications of the Lambert W function to Theoretical Physics Education",
{AIP Conference Proceedings}
(2010) v.~1203, no.~1. 
pp.~1354-1359. 

\bibitem{PI13}
{Pisanova E.S.and Ivanov S.I.} "On the Critical Behavior of the Inverse Susceptibility of a Model of Structural Phase Transitions"
{Bulg. Journ. of Phys.}
(2013) v.~40, no.~2. 
pp.~159-164. 

\bibitem{PI15}
{Pisanova E.S.and Ivanov S.I.}"Non-universal critical properties of the ferromagnetic mean spherical model with long-range interaction",
{Bulg. Chemical Communications}
(2015)
v.~43, no.Spec.Iss.B. pp.~269-274.

\bibitem{G64}
{Griffits R.B.} "A proof that the free energy of a spin system is extensive",
{J. Math. Phys.}
(1964) v.~, no.~9. 
p.~1215. 

\bibitem{F65}
{Fisher M.E.}"Correlation function and coexistence of phases",
{J.Math. Phys.}
(1965} v.~6, . pp.~1643-1653. 

\end{thebibliography}
\end{document}